\documentclass[prd,twocolumn,showpacs,floatfix,nofootinbib,amsmath,amssymb,floatfix]{revtex4}
\usepackage{graphicx,color,dcolumn,booktabs,bm}
\usepackage{mathrsfs}
\usepackage{longtable,lscape}
\usepackage{txfonts}
\usepackage{overpic}
\usepackage{amssymb}
\usepackage{indentfirst}
\usepackage{epsfig}
\usepackage{feynmf}   
\usepackage{epstopdf}   
\usepackage{slashed}  
\usepackage{cases}
\usepackage{color}
\usepackage{multirow}
\usepackage{graphicx,color,dcolumn,booktabs,bm}

\graphicspath{{Figures/}} %

\graphicspath{{Figures/}} %
\usepackage[colorlinks, citecolor=blue,anchorcolor=red,menucolor=red, linkcolor=red,filecolor=red,runcolor=red,urlcolor=blue,frenchlinks=red]{hyperref}

\begin{document}

\title{Constructing $J/\psi$ family with updated data of charmoniumlike $Y$ states}

\author{Jun-Zhang Wang$^{1,2}$}\email{wangjzh2012@lzu.edu.cn}
\author{Dian-Yong Chen$^3$}\email{chendy@seu.edu.cn}
\author{Xiang Liu$^{1,2}$}\email{xiangliu@lzu.edu.cn}
\author{Takayuki Matsuki$^{4,5}$}\email{matsuki@tokyo-kasei.ac.jp}
\affiliation{
$^1$School of Physical Science and Technology, Lanzhou University,
Lanzhou 730000, China\\
$^2$Research Center for Hadron
and CSR Physics, Lanzhou University $\&$ Institute of Modern Physics
of CAS, Lanzhou 730000, China\\
$^3$School of Physics, Southeast University, Nanjing 211189, China\\
$^4$Tokyo Kasei University, 1-18-1 Kaga, Itabashi, Tokyo 173-8602, Japan\\
$^5$Theoretical Research Division, Nishina Center, RIKEN, Wako, Saitama 351-0198, Japan}

\date{\today}

\begin{abstract}
Based on the updated data of charmoniumlike state $Y(4220)$ reported in the hidden-charm 
channels of the $e^+e^-$ annihilation, we propose a $4S$-$3D$ mixing scheme to categorize $Y(4220)$ into the $J/\psi$ family. We find that the present experimental data can support this charmonium assignment to $Y(4220)$.
Thus, $Y(4220)$ plays a role of a scaling point in constructing higher charmonia above 4 GeV.
To further test this scenario, we provide more abundant information on the decay properties of $Y(4220)$, and predict its charmonium partner $\psi(4380)$, whose evidence is found by analyzing the $e^+e^-\to \psi(3686)\pi^+\pi^-$ data from BESIII.
If $Y(4220)$ is indeed a charmonium, we must face how to settle the established charmonium $\psi(4415)$ in the $J/\psi$ family.
In this work, we may introduce a $5S$-$4D$ mixing scheme, and obtain the information of the resonance parameters and partial open-charm decay widths of $\psi(4415)$, which do not contradict the present experimental data.
Additionally, we predict a charmonium partner $\psi(4500)$ of $\psi(4415)$, which can be accessible at future experiments, especially, BESIII and BelleII. The studies presented in this work provide new insights to establish the higher charmonium spectrum.
\end{abstract}

\maketitle

\section{Introduction}

\begin{figure}[htbp]
\begin{center}
\scalebox{0.32}{\includegraphics{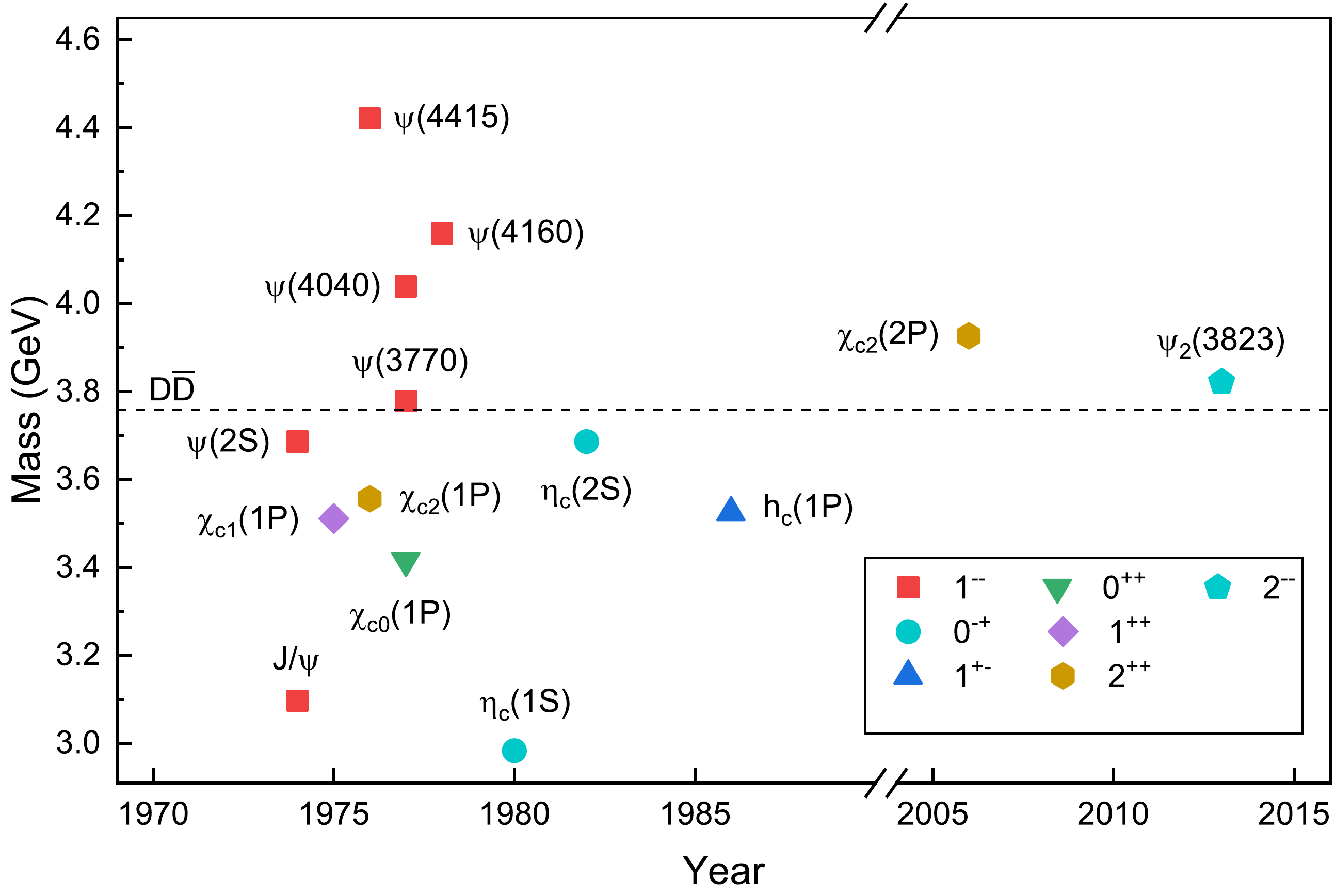}}
\caption{The observed charmonia with the corresponding first observed year \cite{Aubert:1974js,Augustin:1974xw,Abrams:1974yy,Rapidis:1977cv,Goldhaber:1977qn,Brandelik:1978ei,Siegrist:1976br,Tanabashi:2018oca,
Partridge:1980vk,Edwards:1981mq,Baglin:1986yd,Biddick:1977sv,Tanenbaum:1975ef,
Whitaker:1976hb,Uehara:2005qd,Bhardwaj:2013rmw}. Here, the $D\bar{D}$ threshold is also given. \label{ccbar}}
\end{center}
\end{figure}

In 1974, $J/\psi$ particle was discovered by the E598 \cite{Aubert:1974js} Collaboration in the $p+Be\to e^+ +e^-+x$ reaction and the SLAC-SP-017 Collaboration \cite{Augustin:1974xw} in the $e^+e^-$ annihilation at the same time. The observation of $J/\psi$ confirmed the existence of a charm quark predicted by the Glashow-Iliopoulos-Maiani mechanism \cite{Glashow:1970gm}. Since then, a series of charmoniumlike states, $\psi(3686)$ \cite{Abrams:1974yy}, $\psi(3770)$ \cite{Rapidis:1977cv}, $\psi(4040)$ \cite{Goldhaber:1977qn}, $\psi(4160)$ \cite{Brandelik:1978ei}, and $\psi(4415)$ \cite{Siegrist:1976br}, were reported, which construct a main body of the observed charmonium spectrum as shown in Particle Data Group (PDG) \cite{Tanabashi:2018oca}. In Fig. \ref{ccbar}, we collect the corresponding information of the observed charmonia with the year for their first discoveries. It is obvious that the year 1978 is an important
time point since most of charmonia listed in the latest PDG were announced.

Under this experimental background, the Cornell model was proposed by Eichten et al. \cite{Eichten:1974af,Eichten:1978tg}, where
the Cornell potential $V(r)=-k/r+r/a^2$ composed of Coulomb-type and linear potentials, which depicts the interaction between charm and anticharm quarks, was postulated and applied to study the observed charmonia \cite{Eichten:1979ms}. As a successful phenomenological model, the Cornell model can describe the observed charmonia at that time. Inspired by the Cornell model, different potential models were developed by various groups \cite{Barbieri:1975jd,Stanley:1980zm,Carlson:1983rw,Richardson:1978bt,Buchmuller:1980bm,Buchmuller:1980su,Martin:1980rm,Bhanot:1978mj,Quigg:1977dd,
Fulcher:1991dm,Gupta:1993pd,Zeng:1994vj,Ebert:2002pp,Godfrey:1985xj}. Among these, a famous one is the Godfrey-Isgur (GI) model \cite{Godfrey:1985xj}, which has semi-relativistic expression of the kinetic and potential energy terms. The GI model was employed
to quantitatively describe not only meson spectra \cite{Godfrey:1985xj} but also baryon spectra \cite{Capstick:1986bm}.

Let us focus on the charmonium family. As a consequence of studying charmonium spectrum by the Cornell model, the properties of the observed charmonia were decoded, i.e., $J/\psi(3096)$ and $\eta_c(2983)$ are $1S$ states, and $\psi(3686)$, $\psi(4040)$, and $\psi(4415)$ are the first, the second, and the third radial excitations of $J/\psi(3096)$, respectively. $\eta_c(3639)$ is a $2S$ state.
$\psi(3770)$ and $\psi(4160)$ are the ground and the first radial $D$-wave states, respectively. Of course, there exists $2S$ and $1D$ mixing of $\psi(3686)$ and  $\psi(3770)$ as discussed in Ref. \cite{Rosner:2001nm}.
$h_c(3525)$ is a $1P$-wave spin-singlet while $\chi_{c0}(3414)$, $\chi_{c1}(3510)$, and $\chi_{c2}(3556)$ form a $1P$-wave spin-triplet.
This conclusion basically follows the studies \cite{Barnes:2005pb,Radford:2007vd,Ebert:2011jc}. Anyway, we need to keep in mind that the Cornell model is a typical quenched quark models. For higher excitations of the charmonium family, we should be careful to determine their properties only by a quenched quark model.

Since 2003, abundant charmoniumlike $XYZ$ states have been reported by experiments (see Refs. \cite{Liu:2013waa,Chen:2016qju} for a review). As the first observed charmoniumlike state, $X(3872)$ was announced by the Belle Collaboration in the $J/\psi\pi^+\pi^-$ invariant mass spectrum from the $B$ meson decay \cite{Choi:2003ue}. Since the mass of $X(3872)$ is lower than that of $\chi_{c1}(2P)$ state predicted by quenched quark models like the GI model \cite{Godfrey:1985xj} and is close to the threshold of $D\bar{D}^*$ channel, there were extensive discussions of  exotic hadron assignments like the $D\bar{D}^*$ molecular state \cite{Swanson:2003tb,Wong:2003xk} or tetraquark state \cite{Maiani:2004vq,Chen:2010ze}. 
Theorists have not given up the effort to categorize $X(3872)$ into the charmonium family. According to lessons from studying $\Lambda(1405)$ \cite{Kimura:2000sm,Hyodo:2011ur}, $D_{s0}(2317)$ \cite{vanBeveren:2003kd,Dai:2003yg}, and $D_{s1}(2460)$ \cite{vanBeveren:2003jv,Dai:2003yg}, the importance of a coupled-channel effect was realized. 
 If considering the coupled-channel effect, the low mass puzzle of $X(3872)$ can be understood \cite{Kalashnikova:2005ui,Zhang:2009bv,Li:2009zu}, which means that $X(3872)$ as $\chi_{c1}(2P)$ state becomes possible by an unquenched quark model.




The study experience of exploring $X(3872)$ tells us that the coupled-channel effect should be considered seriously, especially for the higher radial and orbital states. When checking the charmonium spectrum, we notice that the $D\bar{D}$ channel is open for $\psi(3770)$. More open-charm decay channels are open for higher states $\psi(4040)$, $\psi(4160)$, and $\psi(4415)$. In 2014, the Lanzhou group once indicated that it is not suitable to assign $\psi(4415)$ as $\psi(4S)$ state. Due to the similarity between  the charmonium and bottomonium families \cite{He:2014xna}, $\psi(4S)$ is roughly predicted to be 4263 MeV by a mass gap estimate, which is also consistent with mass of $\psi(4S)$ predicted by potential models \cite{Dong:1994zj,Li:2009zu} with a color-screening effect. Here, we need to emphasize that there exists some equivalence between the screening potential model and coupled-channel model \cite{Li:2009ad}, which is a reason why our mass value of the $\psi(4S)$ is totally different from quenched quark models.

Frankly speaking, in the past 40 years, the charmonium spectrum above 4.16 GeV was not established, which also reflects how poorly we understand the nonperturbative behavior of Quantum Chromodynamics. This situation stimulates our interest in hunting the evidence of missing higher charmonia by combining with the updated experimental information of charmoniumlike $Y$ states in the $e^+e^-$ annihilations.

\begin{figure}[htbp]
\begin{center}
\scalebox{0.35}{\includegraphics{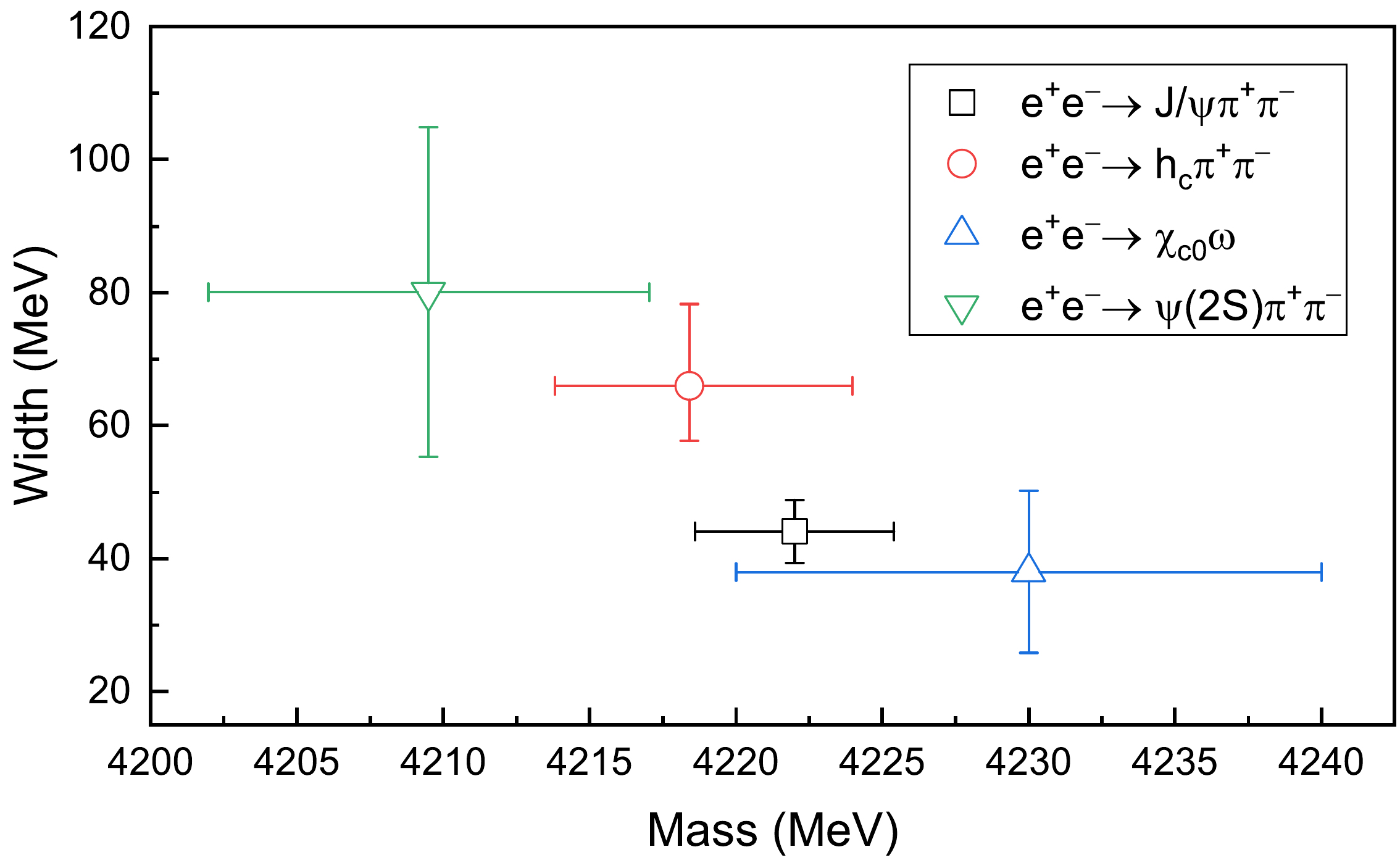}}
\caption{ The measured resonance parameters of $Y(4220)$ in hidden-charm processes $e^+e^-\to J/\psi\pi^+\pi^-$ \cite{Ablikim:2016qzw}, $e^+e^-\to h_c\pi^+\pi^-$ \cite{BESIII:2016adj}, $e^+e^-\to \chi_{c0}\omega$ \cite{Ablikim:2014qwy}, and $e^+e^-\to \psi(3686)\pi^+\pi^-$ \cite{Ablikim:2017oaf}. \label{4220exp}}
\end{center}
\end{figure}

In 2013, BESIII released the measurement of the cross section $e^+e^-\to h_c\pi^+\pi^-$ \cite{Ablikim:2013wzq}, which shows that there may exist a narrow structure around 4.2 GeV \cite{Chang-Zheng:2014haa}. The resonance parameter of this narrow structure is the expected $\psi(4S)$ in Ref. \cite{He:2014xna}. Later, BESIII measured the cross section $e^+e^-\to \omega\chi_{cJ}$ at $\sqrt{s}=4.21-4.42$ GeV, and reported a narrow structure with $m=4230\pm8\pm6$ MeV and $\Gamma=38\pm12\pm2$ MeV \cite{Ablikim:2014qwy}. The Lanzhou group pointed out that this resonance structure is the missing higher charmonium $\psi(4S)$ \cite{Chen:2014sra}. By analyzing updated $e^+e^-\to \pi^+\pi^- \psi(3686)$ data from Belle \cite{Wang:2014hta}, the group again emphasized that the missing $\psi(4S)$ may exist in the $e^+e^-\to \pi^+\pi^- \psi(3686)$ process \cite{Chen:2015bma}, which was confirmed by the BESIII result of $e^+e^-\to \pi^+\pi^- \psi(3686)$ \cite{Ablikim:2017oaf}. In Ref. \cite{Chen:2015bma}, the group also performed a combined fit to the experimental data of $e^+e^-\to \psi(3686)\pi^+\pi^-$ \cite{Wang:2014hta}, $h_c\pi^+\pi^-$ \cite{Ablikim:2013wzq}, and $\chi_{c0}\omega$ \cite{Ablikim:2014qwy}, and found that the narrow structures around 4.2 GeV in different processes can be due to the same state. In 2017, BESIII gave more precise data of $e^+e^-\to J/\psi\pi^+\pi^-$ \cite{Ablikim:2016qzw}, which shows that former super star $Y(4260)$ \cite{Aubert:2005rm} contains two structures $Y(4220)$ and $Y(4330)$. This updated BESIII result announces the end of the era of $Y(4260)$, which lasted 12 years. The updated $e^+e^-\to h_c\pi^+\pi^-$ result from BESIII in 2017 provides the evidence of two structures $Y(4220)$ and $Y(4390)$ existing in the $h_c\pi^+\pi^-$ invariant mass spectrum \cite{BESIII:2016adj}, and the later one is assigned as a $Y(4260)$ partner in the molecular scenario in Ref. \cite{He:2017mbh, Chen:2017abq} . However,  according to these two experimental measurements, the Lanzhou group indicated that only $Y(4220)$ remains while $Y(4330)$ and $Y(4390)$ can be killed by the Fano-like interference effect \cite{Chen:2010nv,Chen:2011kc,Chen:2015bft,Chen:2017uof}, which is from the contributions of two charmonia $\psi(4160)$ and $\psi(4415)$, and the continuum background \cite{Chen:2017uof}. In Fig. \ref{4220exp}, we list the resonance parameters of 4.2 GeV structures in  $e^+e^-\to h_c\pi^+\pi^-$ \cite{BESIII:2016adj}, $e^+e^-\to J/\psi\pi^+\pi^-$ \cite{Ablikim:2016qzw}, $e^+e^-\to \psi(3686)\pi^+\pi^-$ \cite{Ablikim:2017oaf}, and $e^+e^-\to \chi_{c0}\omega$ \cite{Ablikim:2014qwy}.


Hereafter, the structures around 4.2 GeV are collectively referred to as $Y(4220)$.

In this work, we indicate $Y(4220)$ may play a role of the scaling point when constructing the whole charmonium family, especially, higher charmonium above 4 GeV. We need to face several key points: (1) The observed charmonia below 4.2 GeV should be well described even assigning $Y(4220)$ to a charmonium. (2) There must exist a charmonium partner of $Y(4220)$, which is still missing in experiment and whose properties should be predicted. The search for this predicted charmonium partner can be applied to test our scenario. (3) It is also crucial how to settle $\psi(4415)$ in the $J/\psi$ family since $\psi(4415)$ is an established charmonium by different experiments.

To quantitatively illustrate these three key points, we adopt an unquenched potential model to study charmonium mass spectrum, which will be introduced in the next section. Associated with the study of mass spectrum, we further investigate the open-charm decay channels, where the quark pair creation (QPC) model is employed. Thus, their total and partial decay widths can be obtained, which makes us possible to compare with the experimental data and to provide the crucial information to experimental investigation.

Usually, the mixture happens between $nS$-wave and $(n-1)D$-wave states. A typical example in the charmonium family is $\psi(3686)$ and $\psi(3770)$, which can be considered as $2S$-$1D$ wave mixing states. Since masses of $\psi(4S)$ and $\psi(3D)$ are close to each other, the $S$-$D$ wave mixing scheme should be considered. This inspires us to consider $4S$-$3D$ mixing scheme for $Y(4220)$. Our study supports $Y(4220)$ as a $4S$-$3D$ mixing state existing in the $J/\psi$ family since our result is consistent with the present experimental data. In addition, we provide more abundant open-charm decay information, which can be applied to test this explanation of $Y(4220)$.

Besides
putting $Y(4220)$ into the $J/\psi$ family under this $4S$-$3D$ mixing scheme, what is more important is the prediction of its charmonium partner $\psi(4380)$. Under this mixing scheme,
an interesting phenomenon appears, i.e., the predicted $\psi(4380)$ mainly decays into $DD_2^*(2460)$ and has a weak coupling to $D\bar{D}$. Thus, we discuss the possible evidence of $\psi(4380)$ existing in the reported open-charm decay channels \cite{Pakhlova:2008zza,Pakhlova:2007fq}.
According to our studies, we strongly suggest to search for the charmonium partner of $Y(4220)$ via the $DD_1(2430)$, $D^*D^*$, and $DD_2^*(2460)$ channel, which will give a good chance for BESIII and Belle II to observe.

When categorizing $Y(4220)$ in the $J/\psi$ family, we have to face how to settle the well established $\psi(4415)$
in the $J/\psi$ family. In this work, we continue to propose $5S$-$4D$ mixing scheme for $\psi(4415)$ and find that the obtained result strongly suggests this possibility. To give a definite conclusion, we need more precise measurements of $\psi(4415)$ like the resonance parameters, and the partial widths of open-charm and hidden-charm decays. Under this mixing scheme, we naturally predict a charmonium partner of $\psi(4415)$, which is also still missing. In this work, its mass, width and partial decay behavior are obtained. The search for it will be an interesting research issue, and
this $5S$-$4D$ mixing scheme assignment to $\psi(4415)$ can be tested in future.

Stimulated by the existence of $Y(4220)$ in the $e^+e^-\to \psi(3686)\pi^+\pi^-$ process \cite{Ablikim:2017oaf}, we consider whether the predicted charmonium partner of $Y(4220)$ may exist in the present experimental data of $e^+e^-\to \psi(3686)\pi^+\pi^-$. In this work, we reanalyze the data of $e^+e^-\to \psi(3686)\pi^+\pi^-$ by introducing
the Fano interference picture \cite{Chen:2015bft,Chen:2017uof} proposed by us, and find the evidence of the charmonium partner of $Y(4220)$.

This paper is organized as follows. After Introduction, we will give the description of the charmonium spectrum when setting ${Y(4220)}$ as a charmoium state (see Sec. \ref{sec2}) and predicting its charmonium partner. In addition, we discuss how to settle $\psi(4415)$ in the $J/\psi$ family, where $5S$-$4D$ mixing scheme is proposed and the corresponding charmonium partner of $\psi(4415)$ is predicted. In Sec. \ref{sec3}, we continue to analyze the recent $e^+e^-\to \psi(3686)\pi^+\pi^-$ data, and find there should exist the evidence of the predicted charmonium partner of $Y(4220)$.
Finally, the paper ends with a summary in Section \ref{sec4}.

\section{Charmonium spectrum}\label{sec2}

\subsection{A concise introduction of the methods adopted}\label{sec2a}

To provide the description of the charmonium spectrum, in this work, we adopt an unquenched potential model, which has been applied to study heavy-light meson systems \cite{Song:2015nia,Song:2015fha}, kaon family \cite{Pang:2017dlw}, and bottomonium zoo \cite{Wang:2018rjg}.

The interaction between charm quark and anti-charm quark can be expressed by the Hamiltonian \cite{Godfrey:1985xj}
\begin{eqnarray}
\tilde{H}=\left(p^2+m_{c}^2\right)^{1/2}+ \left(p^2+m_{\bar{c}}^2\right)^{1/2} +\tilde{V}_{\mathrm{eff}}\left(\textbf{p},\textbf{r}\right), \label{Eq:Htot}
\end{eqnarray}
where $m_{c}$ and $m_{\bar{c}}$ are the masses of charm quark and anti-charm quark, respectively. $\tilde{V}_{\mathrm{eff}}(\textbf{p},\textbf{r})$ contains a short range $\gamma^\mu\otimes\gamma_\mu$ interaction of one-gluon-exchange and a long range 1$\otimes$1 linear color confining interaction \cite{Godfrey:1985xj}. In the nonrelativistic limit, $\tilde{V}_{\mathrm{eff}}(\textbf{p},\textbf{r})$ can be translated into a familiar nonrelativistic potential $V_{\mathrm{eff}}(r)=H^{\mathrm{conf}}_{q\bar{q}}+H^{\mathrm{hyp}}_{q\bar{q}}+H^{\mathrm{SO}}_{q\bar{q}}$. Here, the first term is a spin-independent potential including the linear confinement and Coulomb-type potential, and the second term denotes the color-hyperfine interaction composed of the tensor and contact terms,
and the third term is from the spin-orbit interaction including the color-magnetic term $H^{\mathrm{SO(cm)}}_{q\bar{q}}$ and the Thomas-precession term
$H^{\mathrm{SO(tp)}}_{q\bar{q}}$ \cite{Godfrey:1985xj}.
There are two aspects reflecting the relativistic corrections \cite{Godfrey:1985xj}, i.e., smearing transformation and momentum-dependent factors. By introducing smearing function
\begin{equation}
\rho \left(\mathbf{r}-\mathbf{r}'\right)=\frac{\sigma^3}{\pi ^{3/2}}\mathrm{exp}\left[-\sigma^2\left(\mathbf{r}-\mathbf{r}'\right)^2\right],
\end{equation}
the confining potential $S(r)=br+c$ and one-gluon exchange potential $G(r)=-4\alpha_s(r)/(3r)$ are smeared out to
\begin{align}
\tilde{S}(r)/\tilde{G}(r)=\int d^3\mathbf{r}'\rho(\mathbf{r}-\mathbf{r}')S(r')/G(r').
\end{align}
This smearing treatment actually takes into account the nonlocality property of interaction between quark and antiquark. Besides, a general relativistic form of the potential should be dependent on momenta of interacting quarks in the center-of-mass system, so a smeared potential $\tilde{V}_i(r)$ could be modified according to
\begin{equation}
{\tilde{V}_i(r)}\to\left(\frac{m_cm_{\bar{c}}}{E_cE_{\bar{c}}}\right)^{1/2+\varepsilon_{i}} {\tilde{V}_i(r)}\left(\frac{m_cm_{\bar{c}}}{E_cE_{\bar{c}}}\right)^{1/2+\varepsilon_{i}},
\end{equation}
where $E_c=(p^2+m_c^2)^{1/2}$ and $E_{\bar{c}}=(p^2+m_{\bar{c}}^2)^{1/2}$ and a parameter $\varepsilon_i$ corresponds to a different type of interaction, such as contact, vector spin-orbit, etc \cite{Godfrey:1985xj}.

In order to include the unquenched effect in this potential model, we need to consider the screening effect, which can be achieved by modifying a linear confining $br+c$ as $$S^{\mathrm{scr}}(r)=\frac{b(1-e^{-\mu r})}{\mu}+c.$$
A similar smearing transformation and momentum-dependent factor for the $S^{\mathrm{scr}}(r)$ are also performed, and the more detailed descriptions of this unquenched potential model can be found in Ref. \cite{Song:2015nia}. To some extent, this unquenched potential model by considering a screening effect is partly equivalent to the coupled channel effect \cite{Li:2009ad,Song:2015nia}. The screening effect has also been supported by unquenched lattice QCD calculations \cite{Bali:2005fu,Namekawa:2011wt}.

When we get the charmonium mass spectrum, the numerical spatial wave functions are obtained by this unquenched potential model, which can be applied to calculate the open-charm decays of the discussed charmonia. To quantitatively study their decay behaviors, we will employ the quark pair creation (QPC) model \cite{Micu:1968mk,LeYaouanc:1977gm}, which is a successful phenomenological method to deal with the Okubo-Zweig-Iizuka (OZI)-allowed strong decays of hadrons.
In the following, we concisely introduce it.

In the QPC model, the transition matrix of the process $A\to B+C$ can be written as
$\langle BC|\mathcal{T}|A \rangle = \delta ^3(\mathbf{P}_B+\mathbf{P}_C)\mathcal{M}^{{M}_{J_{A}}M_{J_{B}}M_{J_{C}}}(\mathbf{P})$, where the transition operator $\mathcal{T}$ describes a quark-antiquark pair creation from the vacuum and reads as
\begin{eqnarray}
\mathcal{T}& = &-3\gamma \sum_{m,i,j}\langle 1m;1~-m|00\rangle\int d \mathbf{p}_3d\mathbf{p}_4\delta ^3 (\mathbf{p}_3+\mathbf{p}_4) \nonumber \\
 && \times \mathcal{Y}_{1m}\left(\frac{\textbf{p}_3-\mathbf{p}_4}{2}\right)\chi _{1,-m}^{34}\phi _{0}^{34}
\left(\omega_{0}^{34}\right)_{ij}b_{3i}^{\dag}(\mathbf{p}_3)d_{4j}^{\dag}(\mathbf{p}_4).
\end{eqnarray}
We introduce a dimensionless constant $\gamma$ depicting the strength of the quark pair creation from the vacuum, which can be fixed by fitting the experimental data. Later, we discuss how to fix it by the present charmonium data.
$\chi_{1,-m}^{34}$ is a spin-triplet state, and $\phi_{0}^{34}$ and $\omega_{0}^{34}$ denote $SU(3)$ flavor and color singlets, respectively. $\mathcal{Y}_{\ell m}(\mathbf{p})\equiv{|\mathbf{p}|^{\ell}}Y_{\ell
m}(\theta_p,\phi_p)$ denotes the $\ell$-th solid harmonic polynomial. By the Jacob-Wick formula \cite{Jacob:1959at}, the helicity amplitudes $\mathcal{M}^{{M}_{J_{A}}M_{J_{B}}M_{J_{C}}}(\mathbf{P})$, which are extracted by the transition matrix element, could be related to the partial wave amplitudes, i.e.,
\begin{eqnarray}
\mathcal{M}^{JL}(A\to BC)&=&\frac{\sqrt{(2L+1)}}{2J_A+1}\sum_{M_{J_B}M_{J_C}}\langle L0;JM_{J_A}|J_AM_{J_A}\rangle \nonumber \\
&&\times \langle J_BM_{J_B};J_CM_{J_C}|{J_A}M_{J_A}\rangle \mathcal{M}^{M_{J_{A}}M_{J_B}M_{J_C}}(\mathbf{P}),\nonumber\\
\end{eqnarray}
where $\mathbf{J}=\mathbf{J}_B+\mathbf{J}_C$, and $\mathbf{L}$ is the orbital angular momentum between final states $B$ and $C$.
The general partial width of the $A\to BC$ reads as
$$\Gamma_{A\to BC}= \pi^2\frac{|\mathbf{P}_B|}{m_A^2}\sum_{J,L}|\mathcal{M}^{JL}(\mathbf{P})|^2.$$
In the above expression, $m_{A}$ is the mass of the initial state $A$.

By this adopted unquenched potential model, we get the numerical spatial wave functions of the involved charmonia and charmed/charm-strange mesons. It can eliminate the parameter dependence of theoretical results compared to the previous calculation \cite{He:2014xna}. In addition, the relevant mass values of involved mesons are taken from the PDG \cite{Tanabashi:2018oca} while the masses of the discussed missing charmonia in this work are from our theoretical calculation.
In our calculation, the constituent quark mass $m_c$, $m_u=m_d$, and $m_s$ are taken as 1.65 GeV, 0.22 GeV and 0.419 GeV, respectively. A parameter $\gamma=5.84$ for $q\bar{q}$ can be extracted by fitting the experimental data as shown in Table \ref{gamma}, where $\chi^2/d.o.f=3.6/3=1.20$ is obtained. And then, the strength for $s\bar{s}$ creation satisfies $\gamma_s=\gamma/\sqrt{3}$ which was suggested in Ref. \cite{LeYaouanc:1977gm}.

\begin{table}[htbp]
\renewcommand\arraystretch{1.2}
\caption{Total widths of observed charmonia above an open charm threshold. $\Gamma_{Expt.}$, $\Gamma_{Th.}$ and $\Gamma_{Error}$ are the experimental data, theoretical width, and fitting error, respectively. The parameter $\gamma=5.84$ is obtained by fitting experimental widths with $\chi^2/d.o.f=1.20$. All results are in units of MeV.   \label{gamma}}
\begin{center}
{\tabcolsep0.08in
\begin{tabular}{ccccc}
\toprule[1pt]
  Charmonium    & $n^{2S+1}L_J$  & $\Gamma_{Expt.}$ \cite{Tanabashi:2018oca} & $\Gamma_{Th.}$ & $\Gamma_{Error}$\\
\midrule[1pt]
$\psi(3770)$ & $1^3D_1$  & $27.2\pm1.0$   & 33.9&6 \\

$\psi(4160)$ & $2^3D_1$  & $70\pm10$  &72.2& 10 \\

$\psi(4040)$ & $3^3S_1$  &$80\pm10$   &66.7& 10  \\

$\chi_{c2}(2P)$&$2^3P_2$ &$24\pm6$  &28.4& 6  \\

\bottomrule[1pt]

\end{tabular}
}
\end{center}
\end{table}

\subsection{Charmonium mass spectrum by scaling $Y(4220)$ as $\psi(4S)$}
\label{sec2b}
In this subsection, the main task is to present the charmonium mass spectrum by the unquenched model when scaling $Y(4220)$ as $\psi(4S)$. Until now, there are fourteen established charmonia \cite{Tanabashi:2018oca} together with $Y(4220)$, which can be employed to limit our potential model parameters. These parameters mainly include the charm quark mass, three screening confinement parameters, and four $\varepsilon_i$ related to the relativistic corrections of momentum factors. By using the following set of parameters,
\begin{eqnarray}
\varepsilon_c=-0.084, & \varepsilon_t=0.012, & \varepsilon_{sov}=-0.053, \varepsilon_{sos}=0.083,\nonumber\\
b=0.2687, & c=-0.3673, &m_c=1.65 \ \mathrm{GeV}, \mu=0.15,
\end{eqnarray}
we find the global aspect of charmonium mass spectrum could be well reproduced. The screening parameter $\mu=0.15$ indicates the unquenched effects may be very important for the charmonium family.  Based on the above parameters, the charmonium spectrum in our unquenched potential model are summarized in Table \ref{spectrum}, where the experimental masses of observed charmonia are also given.  From Table \ref{spectrum}, we can clearly see that the charmonium mass spectrum below 4.2 GeV is well described, especially for the $S$-wave ground states. {In this Table, we identify $Y(4220)$ as our $\psi(4^3S_1)$.} Therefore, the first key point of constructing the whole charmonium spectrum by the updated $Y(4220)$ has been achieved.


\begin{table}[htbp]
\renewcommand\arraystretch{1.3}  
\caption{The charmonium mass spectrum calculated by the unquenched potential model. $Y(4220)$ is identified as $\psi(4^3S_1)$ in our calculation. All results are in units of MeV. Here, we also list the experimental data for comparison with the theoretical results. \label{spectrum}}
\begin{center}
{\tabcolsep0.035in
\begin{tabular}{ccc|ccc}
\toprule[1pt]\toprule[1pt]
    State&     Mass & Expt. \cite{Tanabashi:2018oca} &   State & Mass  & Expt. \cite{Tanabashi:2018oca} \\
\midrule[1pt]
     $\eta_c(1^1S_0)$& 2981 & 2983.9$\pm$0.5   &    $\psi(1^3D_1)$& 3830 & 3778.1$\pm$1.2 \\
     $\psi(1^3S_1)$& 3096 & 3096.9$\pm$0.006   &   $\psi_2(1^3D_2)$& 3848 &  3822.2$\pm$1.2 \\
     $\eta_c(2^1S_0)$& 3642 & 3637.6$\pm$1.2   &   $\psi_3(1^3D_3)$& 3859 & $\cdots$ \\
     $\psi(2^3S_1)$& 3683 & 3686.097$\pm$0.01 &   $\eta_{c2}(2^1D_2)$& 4137 &$\cdots$ \\
     $\eta_c(3^1S_0)$& 4013 &$\cdots$          &   $\psi(2^3D_1)$& 4125 & 4159$\pm$20 \\
     $\psi(3^3S_1)$& 4035 & 4039$\pm$1         &   $\psi_2(2^3D_2)$& 4137 &$\cdots$    \\
     $\eta_c(4^1S_0)$& 4260 &$\cdots$         &    $\psi_3(2^3D_3)$& 4144 &$\cdots$    \\
     $\psi(4^3S_1)$& 4274 &4230$\pm8$     &$\eta_{c2}(3^1D_2)$& 4343 &$\cdots$  \\
     $\eta_c(5^1S_0)$& 4433 &$\cdots$                 &$\psi(3^3D_1)$& 4334 &$\cdots$  \\
     $\psi(5^3S_1)$& 4443 &$\cdots$                   &$\psi_2(3^3D_2)$& 4343 &$\cdots$ \\
     $h_c(1^1P_1)$& 3538 & 3525.38$\pm$0.11     &  $\psi_3(3^3D_3)$& 4348 &$\cdots$\\
     $\chi_{c0}(1^3P_0)$& 3464 & 3414.71$\pm$0.3  & $\eta_{c2}(4^1D_2)$& 4490 &$\cdots$\\
     $\chi_{c1}(1^3P_1)$& 3530 & 3510.67$\pm$0.05   &$\psi(4^3D_1)$& 4484      &$\cdots$\\
     $\chi_{c2}(1^3P_2)$& 3571 & 3556.17$\pm$0.07   &$\psi_2(4^3D_2)$& 4490     &$\cdots$\\
     $h_c(2^1P_1)$& 3933 & $\cdots$                    &   $\psi_3(4^3D_3)$& 4494 &$\cdots$\\
     $\chi_{c0}(2^3P_0)$& 3896 & 3918.4$\pm$1.9   &$h_{c3}(1^1F_3)$& 4074 &$\cdots$\\
     $\chi_{c1}(2^3P_1)$& 3929 &-                 &$\chi_{c2}(1^3F_2)$& 4070 &$\cdots$\\
     $\chi_{c2}(2^3P_2)$& 3952 &3927.2$\pm$2.6   &$\chi_{c3}(1^3F_3)$& 4075 &$\cdots$ \\
     $h_c(3^1P_1)$& 4200 &$\cdots$                &$\chi_{c4}(1^3F_4)$& 4076 &$\cdots$\\
     $\chi_{c0}(3^3P_0)$& 4177 &$\cdots$          &$h_{c3}(2^1F_3)$& 4296 &$\cdots$\\
     $\chi_{c1}(3^3P_1)$& 4197 &$\cdots$          &$\chi_{c2}(2^3F_2)$& 4293 &$\cdots$\\
     $\chi_{c2}(3^3P_2)$& 4213 &$\cdots$          &$\chi_{c3}(2^3F_3)$& 4297 &$\cdots$\\
     $h_c(4^1P_1)$& 4389 &$\cdots$               &$\chi_{c4}(2^3F_4)$& 4298 &$\cdots$ \\
     $\chi_{c0}(4^3P_0)$& 4374 &$\cdots$             &$\eta_{c4}(1^1G_4)$& 4250 &$\cdots$\\
     $\chi_{c1}(4^3P_1)$& 4387 &$\cdots$          &$\psi_3(1^3G_3)$& 4252 &$\cdots$\\
     $\chi_{c2}(4^3P_2)$& 4398 &$\cdots$          &$\psi_4(1^3G_4)$& 4251 &$\cdots$\\
     $\eta_{c2}(1^1D_2)$& 3848 & $\cdots$         &$\psi_5(1^3G_5)$& 4249 &$\cdots$\\

\bottomrule[1pt]\bottomrule[1pt]
\end{tabular}
}
\end{center}
\end{table}

\subsection{$\psi(4S)$ and ${\psi(3D)}$} \label{sec2c}


In this subsection, we firstly discuss the OZI-allowed strong decay behavior  of $\psi(4S)$.
We get the total decay width 27.2 MeV for $\psi(4S)$ when the input mass is chosen as 4274 MeV.  The above results show that treating the charmoniumlike $Y(4220)$ state \cite{Ablikim:2016qzw,BESIII:2016adj,Ablikim:2014qwy}
 as $\psi(4S)$ state is reasonable since the resonance parameter of $Y(4220)$ can be reproduced under the $\psi(4S)$ assignment. Our result also supports the conclusion of the $\psi(4S)$ as a narrow state in Ref. \cite{He:2014xna}.

In the following, we further list the obtained branching ratios of the open-charm decay channels of $\psi(4S)$, i.e.\footnote{{Here, $\psi(4S)\to DD^*$ denotes all the contributions of $\psi(4S)$ decays into two pure neutral states $\phi^{+-}$ and $\phi^{00}$ with a negative $C$ parity, where $\phi^{+-}= \frac{1}{\sqrt{2}}[D^+D^{*-}+\mathcal{C}^{\prime}D^-D^{*+}]$ and $\phi^{00}= \frac{1}{\sqrt{2}}[D^0\bar{D}^{*0}+\mathcal{C}^{\prime}\bar{D}^0D^{*0}]$. Also, $\psi(4S)\to D_sD_s^*$ is an abbreviation of the $\psi(4S)$ decay into a pure neutral system $\phi_{s\bar{s}}^{+-}=\frac{1}{\sqrt{2}}[D_s^+D_s^{*-}+\mathcal{C}^{\prime}D_s^-D_s^{*-}]$. Here, $\mathcal{C}^\prime=+1$ was suggested for the $D\bar{D}^*$ system with a negative $C$ parity (see the discussions in Refs. \cite{Liu:2008fh,Lee:2009hy,Nielsen:2010ij,Artoisenet:2010va,FernandezCarames:2009zz}). We need to emphasize that the results of decay widths are not affected by the convention of $\mathcal{C}^\prime$. When calculating the processes listed in Eqs. (\ref{br3d7})-(\ref{br3d10}), we also need to construct the corresponding pure neutral states by the similar approach. According to the above convention, we obtain the relation of decay amplitude
\begin{eqnarray}
\mathcal{M}_{D\bar{D}^*}=\frac{1}{\sqrt{2}}[\mathcal{M}_{odd}+\mathcal{M}_{even}],
\end{eqnarray}
where $\mathcal{M}_{even}=0$ for charmonia with $J^{PC}=1^{--}$ decay due to the constraint of $C$-parity conservation.
Thus, finally we get $\mathcal{M}_{odd}=\sqrt{2}\mathcal{M}_{D\bar{D}^*}$, where $\mathcal{M}_{D\bar{D}^*}$ can be calculated by the QPC model.
}},
\begin{eqnarray}\label{br3s}
\mathcal{B}[\psi(4S)\to D\bar{D}]&=&9.39 \%,\label{br3s1}\\
\mathcal{B}[\psi(4S)\to D{D}^*]&=&0.347 \%,\label{br3s2}\\
\mathcal{B}[\psi(4S)\to D^*\bar{D}^*]&=&87.7 \%,\label{br3s3}\\
\mathcal{B}[\psi(4S)\to D_s\bar{D}_s]&=&7.13\times 10^{-2} \% ,\label{br3s4}\\
\mathcal{B}[\psi(4S)\to D_s{D}_s^*]&=&2.50 \%,\label{br3s5}\\
\mathcal{B}[\psi(4S)\to D_s^*\bar{D}_s^*]&=&3.38\times 10^{-2} \%\label{br3s6}.
\end{eqnarray}

\begin{figure}[htbp]
\begin{center}
\scalebox{0.42}{\includegraphics{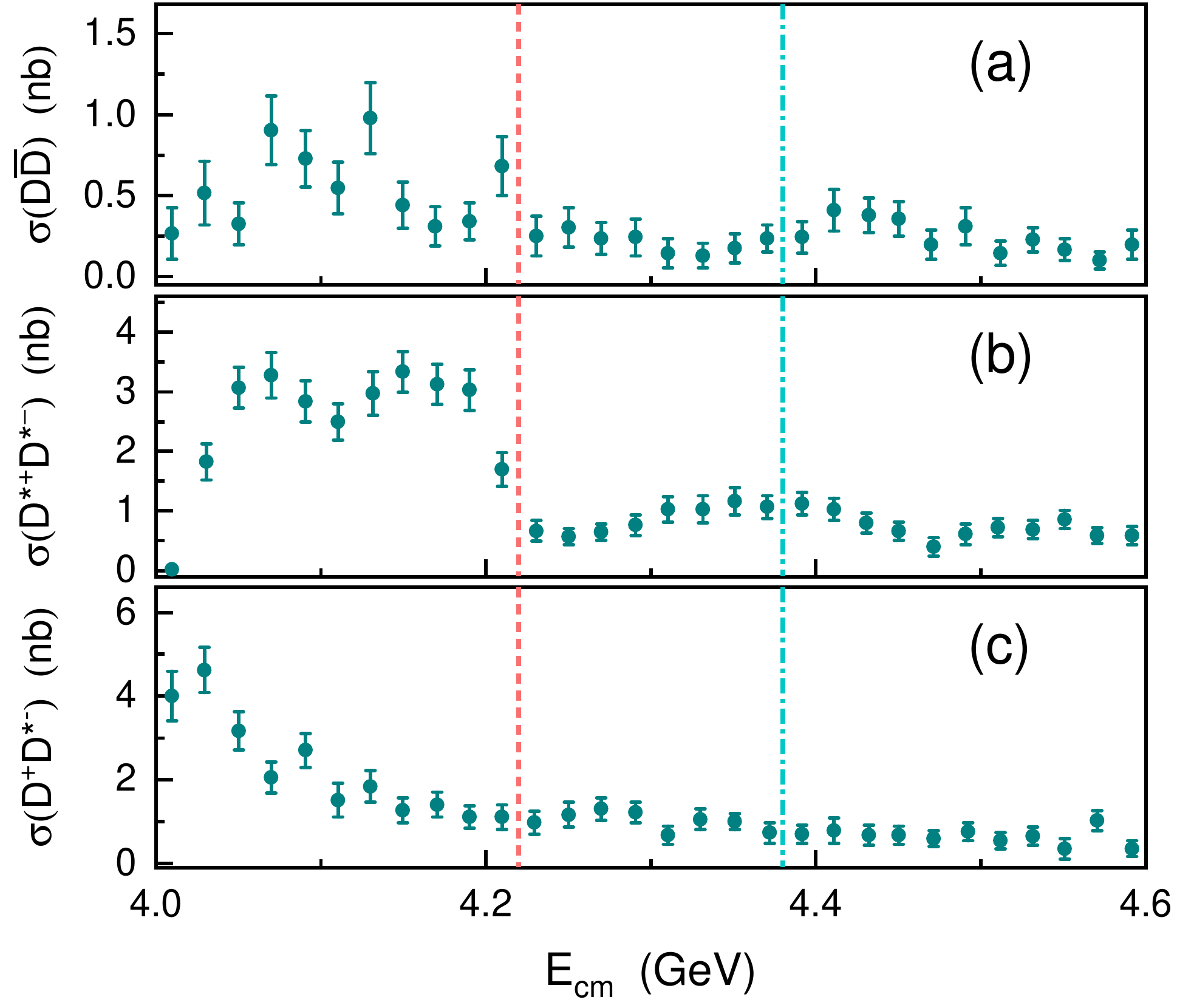}}
\caption{The experimental data of the open charm decay channels from
$e^+e^-$ annihilation. Here, (a) $e^+e^-\to D\bar{D}$
\cite{Pakhlova:2008zza}; (b) $e^+e^-\to D^{*+}{D}^{*-}$
\cite{Abe:2006fj}; (c) $e^+e^-\to D^+ {D}^{*-}$ \cite{Abe:2006fj}.
The red and blue dashed vertical lines correspond to the c.m. energy of 4.22 and 4.38 GeV,
respectively. \label{Opencharm1}}
\end{center}
\end{figure}

For $\psi(4S)$, the main decay modes are composed of six typical open-charm decays just shown in Eqs. (\ref{br3s1})-(\ref{br3s6}). If identifying $Y(4220)$ to be a $\psi(4S)$ state, this $Y(4220)$ structure should be found in the corresponding open-charm decay channels. Especially, our result shows that $D^*\bar{D}^*$ is the dominant decay channel of $\psi(4S)$.
 In Fig. \ref{Opencharm1}, we collect the experimental data of open-charm decay channels from the $e^+e^-$ annihilation, which were released by the Belle Collaboration as early as 2007 \cite{Pakhlova:2008zza,Abe:2006fj}.
There does not exist the evidence of enhancement structures around 4.2 GeV to support this scenario of $Y(4220)$ as $\psi(4S)$\footnote{When carefully checking the $e^+e^-\to D\bar{D}$, we notice one jumping experiment point at $E_{cm}=4.210$ GeV, which may show a possible enhancement. Although this phenomenon should be confirmed by precise measurements, our result indeed gives that $\psi\to D\bar{D}$ is sizable.}.
Here, we should point out that the Belle measurements of the open-charm decay channels are still rough since the bin size of energy is large, which is not enough to provide a definite test of this scenario, especially for a narrow charmonium. Thus, we should wait for more precise data from BESIII and Belle II.

\begin{figure}[htbp]
\begin{center}
\scalebox{0.40}{\includegraphics{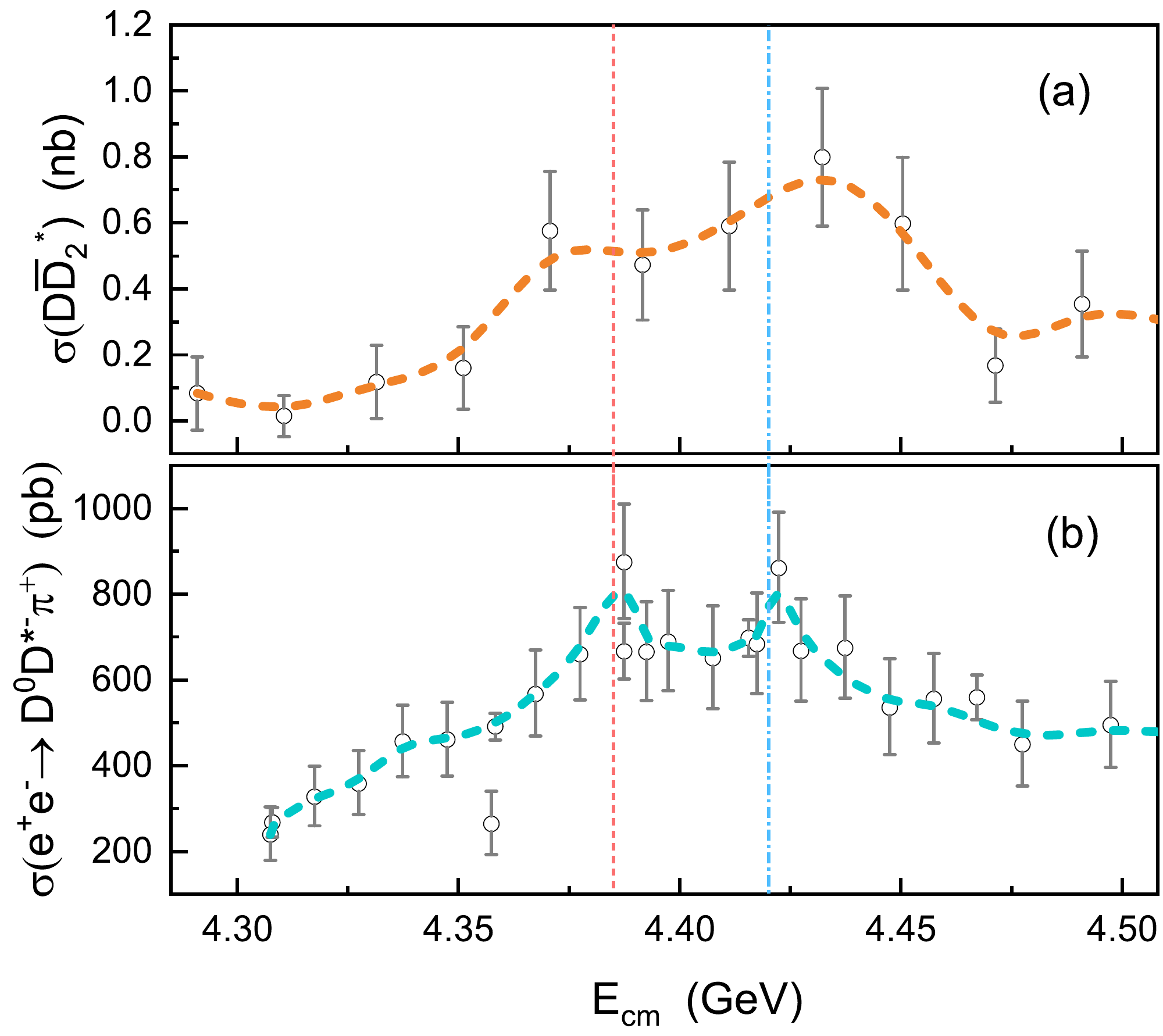}}
\caption{ The experimental data of the open-charm channels $e^+e^-\to D\bar{D}_2^* \to D^0 D^- \pi^+$
\cite{Pakhlova:2007fq} (upper panel) and $e^+e^-\to D^{0}{D}^{*-}\pi^+$
\cite{Ablikim:2018vxx} (lower panel).
The red and blue dashed vertical lines correspond to the c.m. energy of 4.385 and 4.420 GeV, respectively. \label{Opencharm2}}
\end{center}
\end{figure}

In 2018, the BESIII Collaboration released the measurements of the cross section of $e^+e^- \to D^0D^{*-}\pi^+$ \cite{Ablikim:2018vxx}.
In their analysis, the $e^+e^-\to D^*\bar{D}^*\to D^0D^{*-}\pi^+$ contribution was rejected, and the $e^+e^-\to D_2^{*}(2460)^+D^{*-} \to D^0D^{*-}\pi^+$ and $e^+e^-\to {D}_1(2420)^0{D}^{0}\to D^0D^{*-}\pi^+$ are allowed
 \cite{Ablikim:2018vxx}. A clear enhancement around 4.23 GeV was observed in the $D^0D^{*-}\pi^+$ invariant mass spectrum, which hints that this structure may have a strong coupling to the virtual $D_1(2420)^0D^0$ channel. Under our $\psi(4S)$ scenario, this phenomenon can be qualitatively understood. In Sec. \ref{sec2b}, the sreening parameter $\mu$
reflects the importance of a screening effect, which means that the coupled-channel effect plays an important role to modulate a bare state of $\psi(4S)$ due to the partial equivalence  between screening and coupled-channel effects \cite{Li:2009ad}. A bare $\psi(4S)$ state associated with other channels like $D\bar{D}$, $DD^*$, and $D^*\bar{D}^*$ strongly couples to the nearby $D_1(2420)D$ channel, which shifts the original mass of $\psi(4S)$ to the present value 4274 MeV. Here, interaction between $\psi(4S)$ and $D_1(2420)D$ is a typical $S$-wave coupling while
coupling of $\psi(4S)$ with $D\bar{D}$, $DD^*$, and $D^*\bar{D}^*$ occurs via $P$-wave interaction.
Thus, $D_1(2420)D$ is one of the most important channels among the allowed coupled channels for $\psi(4S)$, which results in a chain reaction $e^+e^-\to \psi(4S)\to D^0D^{*-}\pi^+$ via the virtual $D_1(2420)^0D^0$ as revealed by BESIII \cite{Ablikim:2018vxx}.

Although our theoretical results on $\psi(4S)$ are in good agreement with the experimental data of $Y(4220)$,  we cannot fully exclude a possibility of an extotic $Y(4220)$, where a popular one is the charmonium hybrid state assignment to $Y(4220)$ (a detailed discussion can be found in Ref. \cite{Chen:2016qju}).
For the charmonium hybrid, the calculation of QCD sum rule \cite{Zhu:1998sv,Zhu:1999wg} and flux tube model \cite{Close:1994hc} suggest that the decay into two $S$-wave charmed mesons $D^{(*)}\bar{D}^{(*)}$ is suppressed. Instead, the modes of one $S$-wave and one $P$-wave charmed mesons are very important. 
Thus, an experimental study of  the open-charm channel $D^{(*)}\bar{D}^{(*)}$ will provide a crucial test of different assignments to the $Y(4220)$ since a charmonium has open-charm decay behavior different from a charmonium hybrid.

If explaining the charmoniumlike state $Y(4220)$ as a $\psi(4S)$ state, we may expect an existence of its $D$-wave partner $\psi(3D)$ state, which is still missing in experiment.
Our calculation shows that the mass and total width of  $\psi(3D)$ are 4.334 GeV and 28.8 MeV, respectively. Similar to $\psi(4S)$, $\psi(3D)$ is also a narrow charmonium.

The calculated branching ratios of the open-charm decays of $\psi(3D)$ are
\begin{eqnarray}\label{br3s}
\mathcal{B}[\psi(3D)\to D\bar{D}]&=&36.8 \%,\label{br3d1}\\
\mathcal{B}[\psi(3D)\to D{D}^*]&=&4.68 \%,\label{br3d2}\\
\mathcal{B}[\psi(3D)\to D^*\bar{D}^*]&=&32 \%,\label{br3d3}\\
\mathcal{B}[\psi(3D)\to D_s\bar{D}_s]&=& 3.33\times10^{-2} \%,\label{br3d4}\\
\mathcal{B}[\psi(3D)\to D_s{D}_s^*]&=&1.22 \%,\label{br3d5}\\
\mathcal{B}[\psi(3D)\to D_s^*\bar{D}_s^*]&=& 0.583\%,\label{br3d6}\\
\mathcal{B}[\psi(3D)\to DD_1(2420)]&=&7.65 \%,\label{br3d7}\\
\mathcal{B}[\psi(3D)\to DD_1(2430)]&=&16.4 \%,\label{br3d8}\\
\mathcal{B}[\psi(3D)\to DD_2^*(2460)]&=&0.235 \%,\label{br3d9}\\
\mathcal{B}[\psi(3D)\to D^*D_0(2400)]&=&0.408 \% \label{br3d10}.
\end{eqnarray}
Thus, $D\bar D$ channel is the dominant decay mode of the $\psi(3D)$ state. The Belle data of $e^+e^-\to D\bar{D}$, however, does not show the evidence of $\psi(3D)$ as presented in Fig. \ref{Opencharm1} (a).

We try to find the evidence of $\psi(3D)$ in the reported data of charmoniumlike states and notice the famous $Y(4360)$ from the $e^+e^- \to \psi(3686)\pi^+\pi^-$ \cite{Tanabashi:2018oca}.  The mass of $Y(4360)$ is close to that of $\psi(3D)$, but the width of $Y(4360)$ is broader than the predicted $\psi(3D)$. This deviation should be faced when treating $Y(4360)$ as $\psi(3D)$. 
In addition, the $e^+e^-$ annihilation decays of $D$-wave vector quarkonium states are generally one to three orders of magnitude smaller than those of corresponding $S$-wave states \cite{Wang:2018rjg}. Thus, it is not an easy task to observe this  $3D$ state through the hidden charm decay channels from the electron-positron annihilation.


When further checking the early data of the open-charm process $e^+e^- \to D\bar{D}^*_2 \to D^0D^-\pi^+$ in Fig. \ref{Opencharm2} (a), a suspicious signal at 4.37 GeV is found.
We may consider whether this enhancement structure is the predicted $\psi(3D)$.
However, our result indicates that $\psi(3D) \to DD^*_2(2460)$ has a tiny partial width (67.6 keV). It is obvious that this structure in $e^+e^- \to D\bar{D}^*_2 \to D^0D^-\pi^+$ cannot explain a $\psi(3D)$ state.
To understand this puzzling phenomenon, we need a new idea.

As mentioned in Introduction, the established charmonium states $\psi(3686)$ and $\psi(3770)$ are admixtures with a small $S$-$D$ mixing angle rather than a  pure $S$-wave or $D$-wave state \cite{Rosner:2001nm}.
This lesson tells us that $4S$-$3D$ mixing scheme should be considered, which may shed light on the above puzzling phenomenon. In the next subsection, we pay more attention to this issue.

\subsection{$4S$-$3D$ mixing scheme}\label{sec2d}

\begin{figure}[htbp]
\begin{center}
\scalebox{0.32}{\includegraphics{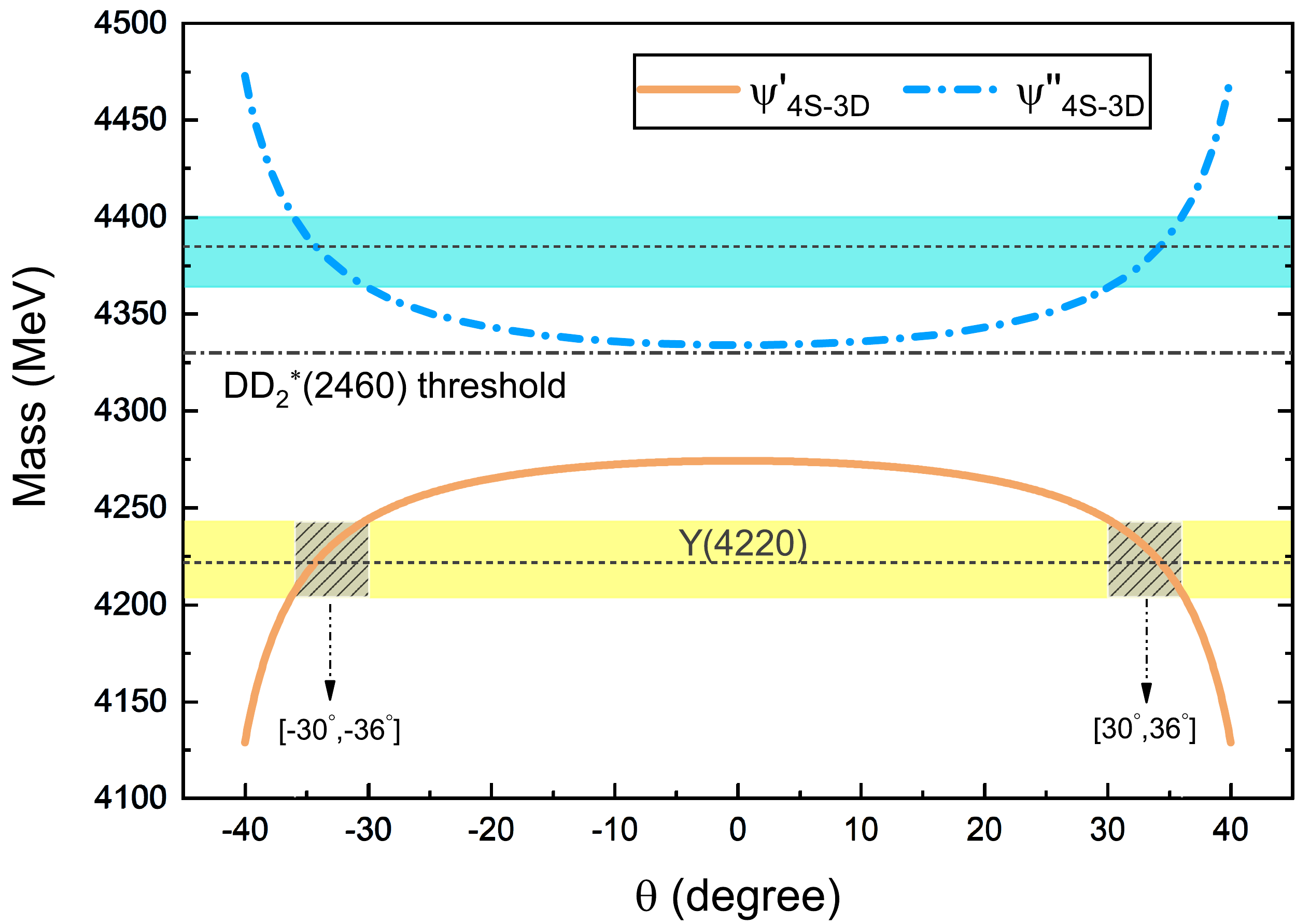}}
\caption{ The masses of $\psi^{\prime}_{4S-3D}$ and $\psi^{\prime\prime}_{4S-3D}$ with dependence on the $4S$-$3D$ mixing angle $\theta$. The yellow and cyan bands denote the measured mass range of $Y(4220)$ and the predicted mass range of $\psi(4380)$, respectively, where the dashed horizontal lines in bands corresponds to each average value. The two shaded regions represent the mixing angle interval, in which the theoretical results of $\psi^{\prime}_{4S-3D}$ meet the measurements of $Y(4220)$. \label{4s3dmass}}
\end{center}
\end{figure}

In this subsection, we discuss the $4S$-$3D$ mixing scheme. Under this framework, we introduce
\begin{eqnarray}
\left( \begin{array}{c}  |\psi_{4S-3D}^\prime\rangle\\  |\psi_{4S-3D}^{\prime\prime}\rangle\end{array} \right) =
\left( \begin{array}{cc} \cos{\theta} & \sin{\theta} \\
                         -\sin{\theta} & \cos{\theta} \end{array} \right)
\left( \begin{array}{c} |4^3S_1\rangle\\ |3^3D_1\rangle \end{array} \right)
\end{eqnarray}
to describe the $4S$-$3D$ mixing. Here, $\theta$ denotes the mixing angle.
Then, the mass eigenvalues of $\psi_{4S-3D}^\prime$ and $\psi_{4S-3D}^{\prime\prime}$ are determined by the masses of two basis vectors $m_{4S}$, $m_{3D}$ and the mixing angle $\theta$, i.e.,
\begin{eqnarray}
m_{\psi_{4S-3D}^\prime}^2=\frac{1}{2}\left(m_{4S}^2+m_{3D}^2-\sqrt{(m_{3D}^2-m_{4S}^2)^2\sec^22\theta}\right)\label{sdmassf1},\\
m_{\psi_{4S-3D}^{\prime\prime}}^2=\frac{1}{2}\left(m_{4S}^2+m_{3D}^2+\sqrt{(m_{3D}^2-m_{4S}^2)^2\sec^22\theta}\right)\label{sdmassf2}.
\end{eqnarray}
As shown in Table \ref{spectrum}, the masses of pure $4S$ and $3D$ $c\bar{c}$ states are obtained by our unquenched potential model. Thus, we take the mass $m_{4S}=4274$ MeV and $m_{3D}=4334$ MeV as input, and present the dependence of $m_{\psi_{4S-3D}^\prime}$ and $m_{\psi_{4S-3D}^{\prime\prime}}$ on $\theta$ (see Fig. \ref{4s3dmass})\footnote{If checking former work of the GI model, we find that the treatment to mixing scheme under the framework of the GI model is not good enough. For example, $K_1(1270)$ and $K_1(1400)$ are mixing states of $1^1P_1$ and $1^3P_1$ states. In Ref. \cite{Godfrey:1986wj}, the mixing angle $\theta_{K_1}$ determined by the spin-orbit interaction of the GI model is $-5^{\circ}$, which is far smaller than the mixing angle extracted by the experimental information \cite{Cheng:2011pb}. Considering this situation, in this work we discuss the $S$-$D$ mixing by a phenomenological approach without adopting the direct calculation by the GI model. A comprehensive study of mixing phenomena in charmonium family is a very interesting research topic, which should be seriously investigated in future work.}. This figure shows that $m_{\psi_{4S-3D}^\prime}$ ($m_{\psi_{4S-3D}^{\prime\prime}}$) becomes lower (higher) than $m_{4S}$ ($m_{3D}$) when increasing the absolute value of $\theta$.

Focusing on the interesting $Y(4220)$, we discuss  the treatment of  $Y(4220)$ as $\psi_{4S-3D}^\prime$. According to the experimental results for $Y(4220)$ \cite{Ablikim:2016qzw,BESIII:2016adj,Wang:2014hta,Ablikim:2014qwy,Ablikim:2018vxx}, we set the mass range of $Y(4220)$ to be $4204 \sim 4243$ MeV, which is lower than the mass of $\psi(4S)$. In fact, this mass difference between $Y(4220)$ and pure $\psi(4S)$ is also a main motivation to stimulate us to introduce  the $4S$-$3D$ mixing scheme.

Using the mass range of $Y(4220)$, we may predict the mass range ($4364\sim4400$ MeV) for $\psi_{4S-3D}^{\prime\prime}$ given in Fig. \ref{4s3dmass}. Since its central value is 4384 MeV, we tentatively name this $\psi_{4S-3D}^{\prime\prime}$ state as $\psi(4380)$ in the following discussion, which is nothing but the partner of the discussed $Y(4220)$. The corresponding mixing angle $\theta=\pm (30^{\circ}\scriptsize{\sim}36^{\circ})$  is obtained.

\begin{figure}[htbp]
\begin{center}
\scalebox{0.36}{\includegraphics{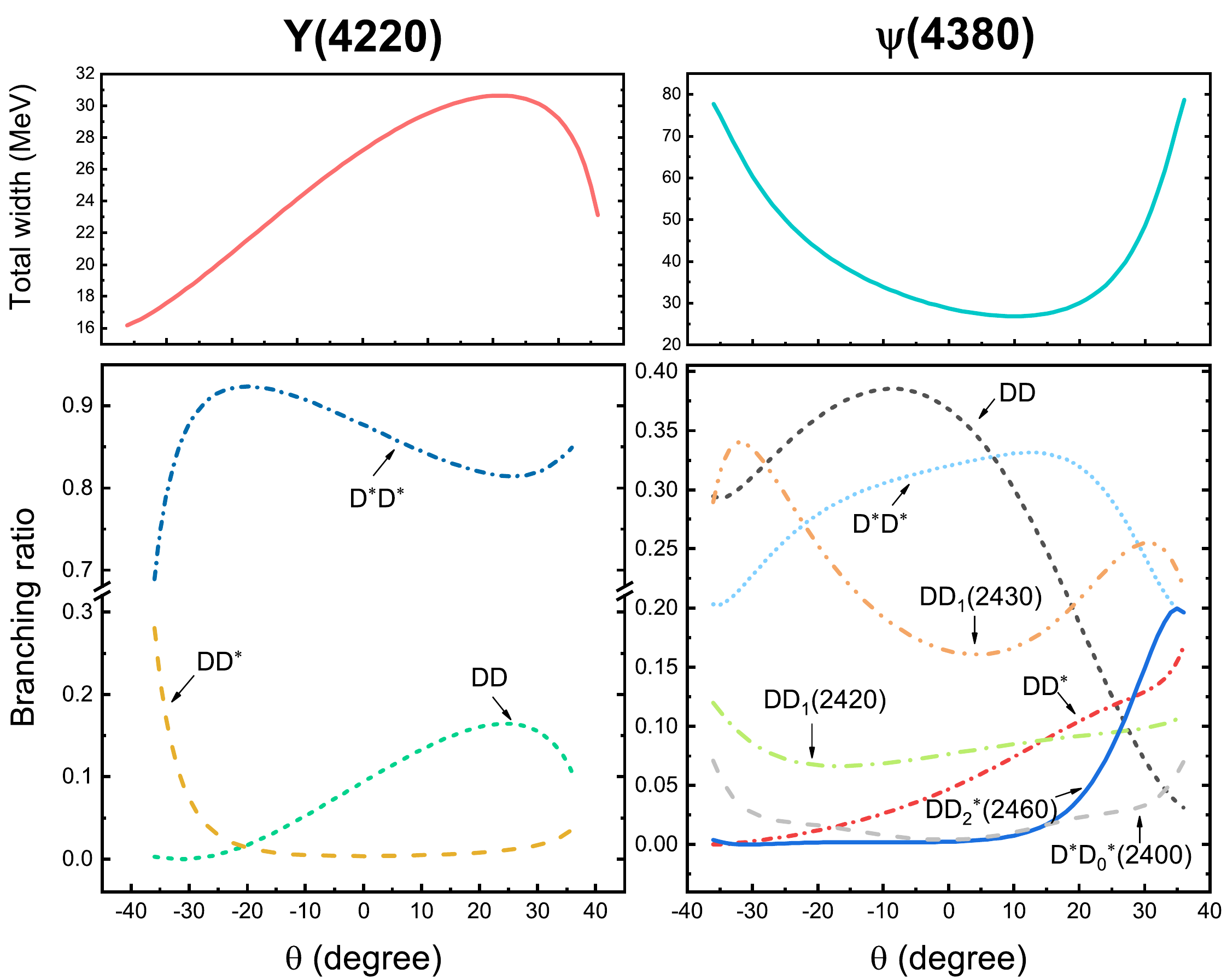}}
\caption{ The open-charm decay behaviors of $Y(4220)$ (left panel) and $\psi(4380)$ (right panel) as a function of the $4S$-$3D$ mixing angle $\theta$. \label{4s3ddecay}}
\end{center}
\end{figure}

In the $4S$-$3D$ mixing scheme, we need to illustrate the decay properties of $Y(4220)$, and further give the decay behaviors of its partner $\psi(4380)$, which are collected in Fig. \ref{4s3ddecay}. We notice that the decay behavior of $Y(4220)$ under this $4S$-$3D$ mixing scheme is similar to that of $Y(4220)$ as a pure $\psi(4S)$ state when taking a positive mixing angle. That is, the obtained total decay width is 26.0 MeV, and the obtained partial widths of the allowed strong decays of $Y(4220)$
are listed in Table \ref{sdmixing}, where
a typical $\theta=+34^\circ$, which corresponds to an average measured mass of 4222 MeV for $Y(4220)$, is taken. When taking a negative mixing angle, the decay property of $Y(4220)$ under this $4S$-$3D$ mixing scheme is different from that of $Y(4220)$ as a pure $\psi(4S)$ state, where the total width of $Y(4220)$ as a mixture of $4S$ and $3D$ states becomes smaller and the branching ratio of the $DD^*$ mode is  larger than that of the $DD$ mode.
Considering the above two cases, we suggest to adopt a positive mixing angle $\theta= (30^{\circ}\scriptsize{\sim}36^{\circ})$ in the following discussion.



Next, we investigate the decay behaviors of $\psi(4380)$ with the running of a mixing angle in Fig. \ref{4s3ddecay}. To our surprise, two main conclusions can be made for the above mixing scheme:

\begin{enumerate}

\item{The total width of $\psi(4380)$ has a significant enhancement, which shows that $\psi(4380)$ should be a broad state since its total width is nearly three times larger than that of a pure $\psi(3D)$ state (28.8 MeV). This conclusion can be understood as follows. Since the phase space from the decays of $\psi(4380)$ into $P$-wave and S-wave charmed mesons is larger than that of a pure $\psi(3D)$ state, the channels of $\psi(4380)$ have large contributions to the total decay width of $\psi(4380)$.}

\item{The dominant decay channels of $\psi(4380)$ are $DD_1(2430)$, $D^*D^*$, and $DD_2^*(2460)$, especially sizable enhancement of $\mathcal{B}(\psi(4380)\to DD_2^*(2460))$. Additionally, the contribution of the $DD$ mode to the total width becomes unimportant. Thus, the decay behavior of $\psi(4380)$ is totally different from a pure $\psi(3D)$ state. This result can be due to the change of the spatial wave function of $\psi(4380)$ obtained in the $4S$-$3D$ mixing scheme.}

\end{enumerate}
The concrete values reflecting the decay behaviors of $\psi(4380)$ are shown in Table \ref{sdmixing}.

The above information indicates that the potential enhancement structure around 4.37 GeV existing in the $e^+e^- \to D\bar{D}^*_2 \to D^0D^-\pi^+$ process (see Fig. \ref{Opencharm2} (a)) can be the predicted $\psi(4380)$ because the $\psi(4380)\to DD_2^*(2460)$ is one of the dominant decay channels. Due to the tiny branching ratio of $\psi(4380)\to D\bar{D}$, there should not exist any signal of $\psi(4380)$ in the reported $e^+e^-\to D\bar{D}$ data as shown in Fig. \ref{Opencharm1} (a).
In addition, another experimental evidence of $\psi(4380)$ associated with the open charm process comes from the latest measurement of $e^+e^- \to D^0D^{*-}\pi^+$ \cite{Ablikim:2018vxx}, where a broad enhancement around 4.40 GeV is visible except for the observation of $Y(4220)$. The BESIII Collaboration indicated the broad structure may be from the contributions of $\psi(4415)$ and other resonances \cite{Ablikim:2018vxx}. Depicting the experimental data shown in Fig. \ref{Opencharm2} (b), we can see two clear enhancements near 4.38 GeV and 4.42 GeV.  The former one implies an unknown resonance and the latter one can be related to the established $\psi(4415)$. From our theoretical point of view, the $\psi(4380)$ state mainly decays to $D^0D^{*-}\pi^+$ through the most dominant mode $D^0D_1(2430)^0$ and an important channel $D^0D_1(2420)^0$. Therefore, the recent BESIII experimental data can support our prediction of a missing charmonium $\psi(4380)$.
In general, the existence of $\psi(4380)$ predicted in the present work does not contradict the announced experimental results. We strongly suggest that experimentalists carry out precise measurements on the $e^+e^- \to D^0D^{*-}\pi^+$ and $e^+e^- \to D^0D^{-}\pi^+$ processes, which will provide a crucial test to our predictions. This is an excellent opportunity for the upgraded Belle II and the running BESIII.

We also calculate the $e^+e^-$ annihilation width of $Y(4220)$ and $\psi(4380)$
by the formula with the first-order QCD radiative corrections given in Refs. \cite{Kwong:1987ak,Bradley:1980eh}, i.e.,
\begin{eqnarray}
&&\Gamma(Y(4220)\to e^+e^-) \nonumber \\
&&=\frac{16\pi e_c^2\alpha_0^2\mathcal{C}}{m_{Y(4200)}^2} \left|\cos \theta R_{4S}(0)+\frac{5\sqrt2}{m_{Y(4220)}^2}\sin \theta R_{3D}^{\prime\prime}(0)\right|^2,  \\
&&\Gamma(\psi(4380)\to e^+e^-) \nonumber \\
&&=\frac{16\pi e_c^2\alpha_0^2\mathcal{C}}{m_{\psi(4380)}^2} \left|\frac{5\sqrt2}{m_{\psi(4380)}^2}\cos \theta R_{3D}^{\prime\prime}(0)-\sin \theta R_{4S}(0)\right|^2,
\end{eqnarray}
where $e_c=2/3$ and $\alpha_0=1/137$ are the charm quark charge and fine structure constant, respectively, $R_{4S}(0)$ is the radial $4S$-wave function at the origin, $R^{\prime\prime}_{3D}(0)$ is the second derivative of the radial $3D$-wave function at the origin, and $\mathcal{C}=(1-\frac{16 \alpha_s}{3\pi})$ corresponds to the first-order QCD radiative correction with $\alpha_s=0.26$ \cite{Kwong:1987ak}.
With the above expression and taking our obtained numerical spatial wave functions as input, we estimate the typical widths of 0.290 keV and 0.257 keV for $Y(4220) \to e^+e^-$ and $\psi(4380) \to e^+e^-$, respectively, by setting $\theta=+34^{\circ}$.

In Ref. \cite{Chen:2015bma}, we once suggested the search for $Y(4220)$ via the hidden-charm process $e^+e^- \to \psi(3686)\pi^+\pi^-$. Fortunately, the recent BESIII's analysis of the $e^+e^- \to \psi(3686)\pi^+\pi^-$ indeed shows the existence of a signal $Y(4220)$ \cite{Ablikim:2017oaf}. Stimulated by this, it is interesting to search for the evidence of the predicted $\psi(4380)$ by this hidden-charm decay process. In Sec. \ref{sec3}, we concretely discuss whether the present data of $e^+e^- \to \psi(3686)\pi^+\pi^-$ from BESIII \cite{Ablikim:2017oaf} contains the signal of the predicted $\psi(4380)$.

\begin{table*}[htbp]
\renewcommand\arraystretch{1.3}
\caption{Partial widths of open-charm strong decays for the charmonium $\psi(4220)$, $\psi(4380)$, $\psi(4415)$, and $\psi(4500)$ in units of MeV. Here, mixing angles for $4S$-$3D$ wave and  $5S$-$4D$ wave admixtures are taken as $\pm34^{\circ}$ and $\pm30^{\circ}$, respectively.  \label{sdmixing}}
\begin{center}
{\tabcolsep0.08in
\begin{tabular}{lcccc|cccc}
\toprule[1pt]\toprule[1pt]
 & \multicolumn{4}{c}{Scheme I with positive angle } & \multicolumn{4}{c}{Scheme II with negative angle} \\
 \cmidrule[1pt]{2-3}\cmidrule[1pt]{4-9}
Channels            &  $\psi(4220)$    &$\psi(4380)$   & $\psi(4415)$ & $\psi(4500)$   &  $\psi(4220)$    &$\psi(4380)$   & $\psi(4415)$ & $\psi(4500)$  \\
\midrule[1pt]
$DD$                     &3.29                &2.74                & 4.14   &   2.01 & 0.0184     &21.3   & $7.93\times 10^{-3}$  &   10.4  \\
 $DD^*$                   &0.723                &10.1            & $5.00\times10^{-3}$      &   3.72& 3.04       &0.0338   &  0.0538            &   0.112  \\
$DD_1(2420)$             &$\cdots$                &7.14             & 2.94           &   4.24 &$\cdots$         &7.77   & 2.92           &   4.60    \\
 $DD_1(2430)$             &$\cdots$                &16.5            & 3.31           &   6.26 &$\cdots$         &23.7   & 1.76           &   7.35  \\
 $DD_2^*(2460)$           &$\cdots$                &13.5          & 0.109   &  4.05   & $\cdots$  &0.102  & 0.943     &  $2.13\times10^{-3}$     \\
 $DD(2550)$               &$\cdots$            & $\cdots$           & 0.0566          &  0.853& $\cdots$      & $\cdots$ & 0.0455         &  6.25         \\
 $DD^*(2600)$               &$\cdots$            & $\cdots$       & $\cdots$     & $2.77\times10^{-4}$ & $\cdots$ & $\cdots$ & $\cdots$      & 0.0674         \\
 $D^*D^*$                 &21.8                &14.0                &  8.42             &  6.50 & 12.9        &14.8      &  5.65         &  7.02     \\
 $D^*D_0^*(2400)$         &$\cdots$                &3.65            & 0.685           &   3.48 & $\cdots$     &3.65      & 0.685         &   3.48   \\
 $D^*D_1(2420)$             &$\cdots$                &$\cdots$            & $\cdots$  &   0.284& $\cdots$     &$\cdots$  & $\cdots$       &   0.305   \\
 $D^*D_1(2430)$             &$\cdots$                &$\cdots$         & $\cdots$     &   0.259 & $\cdots$  &$\cdots$    & $\cdots$      &   0.199   \\
 $D^*D_2^*(2460)$           &$\cdots$                &$\cdots$        & $\cdots$      & 0.466  & $\cdots$   &$\cdots$    & $\cdots$      & 0.737     \\
 $D_sD_s$                 &0.144       &0.0246    &  $2.33\times 10^{-3}$    & 0.0113 & $7.89\times10^{-4}$  &0.169 &  $1.03\times 10^{-4}$   & 0.0586       \\
 $D_sD_s^*$               &0.0486           &0.610      & 0.0625    &  0.244& 0.605   &$1.10\times10^{-4}$     & 0.448       &    $3.61\times10^{-3}$   \\
 $D_s^*D_s^*$             &$\cdots$            &0.330               & 0.222          &  0.124 & $\cdots$       &0.938     & 0.119          &  0.409   \\

\midrule[1pt]
 Total					   &26.0 MeV	      &68.6 MeV              & 19.9 MeV          & 	32.6 MeV & 16.5 MeV	 &72.5 MeV  & 12.6 MeV  & 41.1 MeV		    \\

\bottomrule[1pt]\bottomrule[1pt]
\end{tabular}
}
\end{center}
\end{table*}

\subsection{Settlement of $\mathbf{\psi(4415)}$ in the $J/\psi$ family}\label{sec2e}

Although $\psi(4415)$ was firstly reported in 1976 \cite{Siegrist:1976br}, its inner structure of  $\psi(4415)$ is still waiting for being revealed. When categorizing $Y(4220)$ into the $J/\psi$ family, we must face how to settle ${\psi(4415)}$ in the $J/\psi$ family, which is one of the main tasks in this work.

\begin{figure}[htbp]
\begin{center}
\scalebox{0.32}{\includegraphics{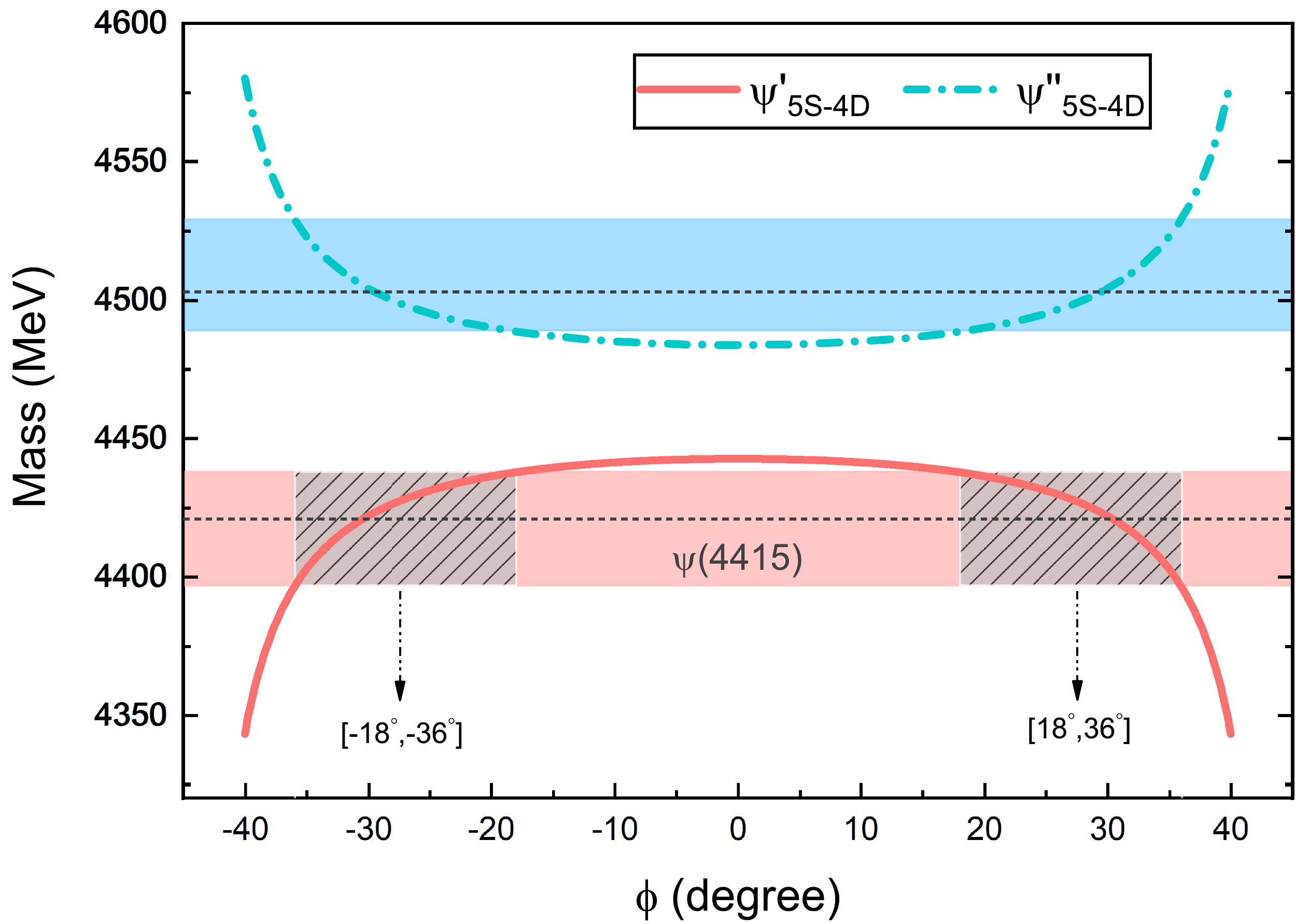}}
\caption{ The masses of $\psi^{\prime}_{5S-4D}$ and $\psi^{\prime\prime}_{5S-4D}$ with dependence on the $5S$-$4D$ mixing angle $\phi$. The red and blue bands denote the measured mass range of $\psi(4415)$ and the predicted mass range of $\psi(4500)$, respectively, where the dashed horizontal lines correspond to each average value. The two shaded regions represent the mixing angle interval, in which the theoretical results of $\psi^{\prime}_{5S-4D}$ meet the measurements of $\psi(4415)$. \label{5s4dmass}}
\end{center}
\end{figure}

The mass spectrum result in Table \ref{spectrum} shows that the mass of ${\psi(4415)}$ is close to that of $\psi(5S)$. Thus, we propose $5S$-$4D$ mixing scheme to study ${\psi(4415)}$, which also borrows the idea when dealing with $Y(4220)$ in Sec. \ref{sec2d}. To depict this mixing scheme, we have an expression
\begin{eqnarray}
\left( \begin{array}{c}  |\psi_{5S-4D}^\prime\rangle\\  |\psi_{5S-4D}^{\prime\prime}\rangle\end{array} \right) =
\left( \begin{array}{cc} \cos{\phi} & \sin{\phi} \\
                         -\sin{\phi} & \cos{\phi} \end{array} \right)
\left( \begin{array}{c} |5^3S_1\rangle\\ |4^3D_1\rangle \end{array} \right),
\end{eqnarray}
where $\phi$ is a mixing angle. The masses of $\psi(5S)$ and $\psi(4D)$ are taken from our calculations listed in Table \ref{spectrum}. Then, the dependence of the masses of $\psi_{5S-4D}^\prime$ and $\psi_{5S-4D}^{\prime\prime}$ on $\phi$ is given in Fig. \ref{5s4dmass}.

In the following, we discuss the possibility of $\psi(4415)$ as $\psi_{5S-4D}^\prime$ state.
In our study, the mass range of $\psi(4415)$ is from PDG \cite{Tanabashi:2018oca}, i.e., $m_{\psi(4415)}=4397\sim4438$ MeV.
Thus, we obtain the mixing angle $\phi=\pm(18\scriptsize{\sim}36)^\circ$ and the predicted mass of the $\psi_{5S-4D}^{\prime\prime}$ state to be $4489\sim4529$ MeV as shown in Fig. \ref{5s4dmass}. The typical value $\phi=\pm 30^\circ$ and $m_{\psi_{5S-4D}^{\prime\prime}}=4503$ MeV directly correspond to a central value of mass of $\psi(4415)$.
This $\psi_{5S-4D}^{\prime\prime}$ state is named as $\psi(4500)$ for convenience of the following discussion.

\begin{figure}[htbp]
\begin{center}
\scalebox{0.35}{\includegraphics{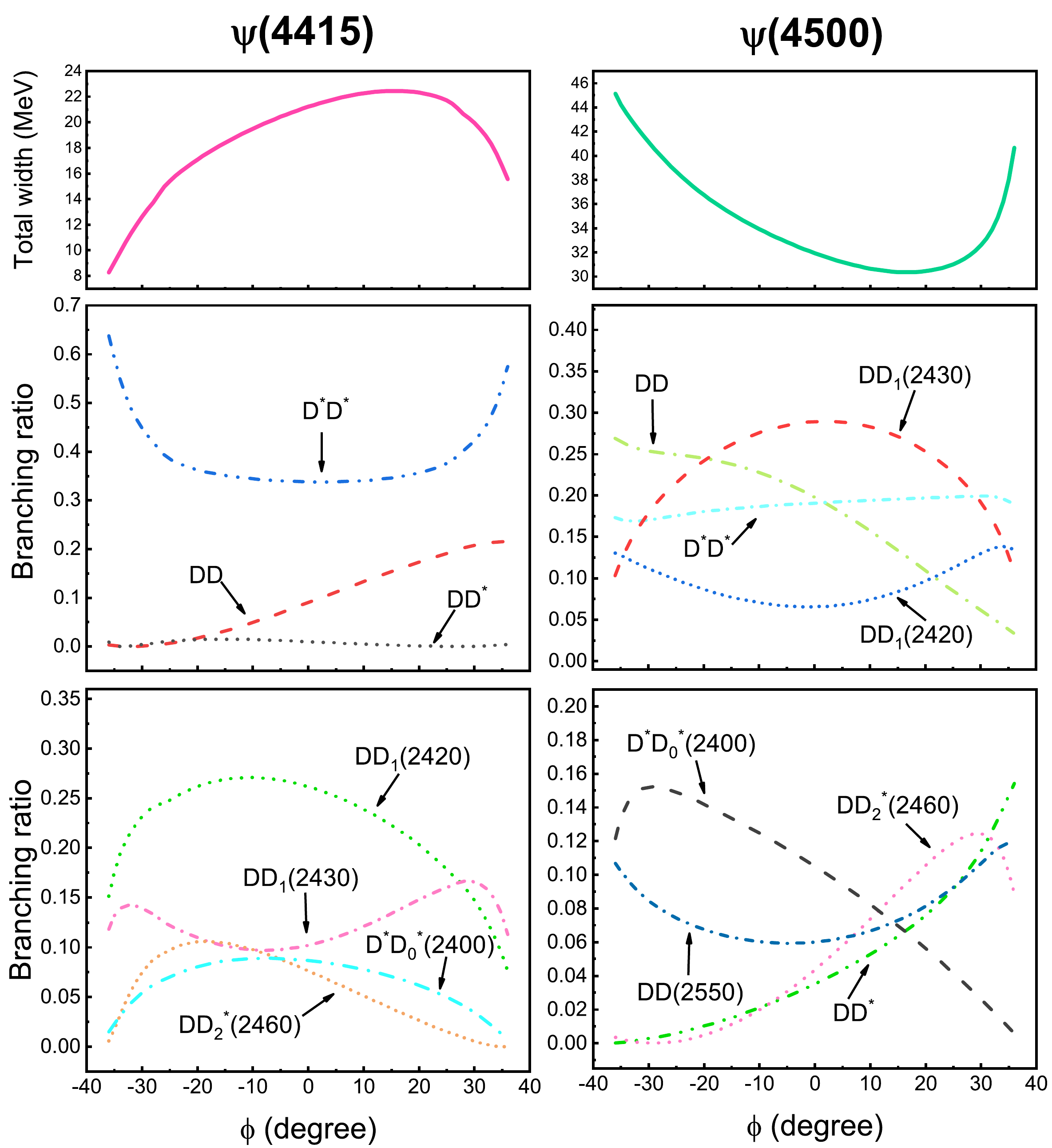}}
\caption{ The open-charm decay behaviours of $\psi(4415)$ (left panel) and $\psi(4500)$ (right panel) as a function of the $5S$-$4D$ mixing angle $\phi$. \label{5s4ddecay}}
\end{center}
\end{figure}

\begin{figure}[htbp]
\begin{center}
\scalebox{0.30}{\includegraphics{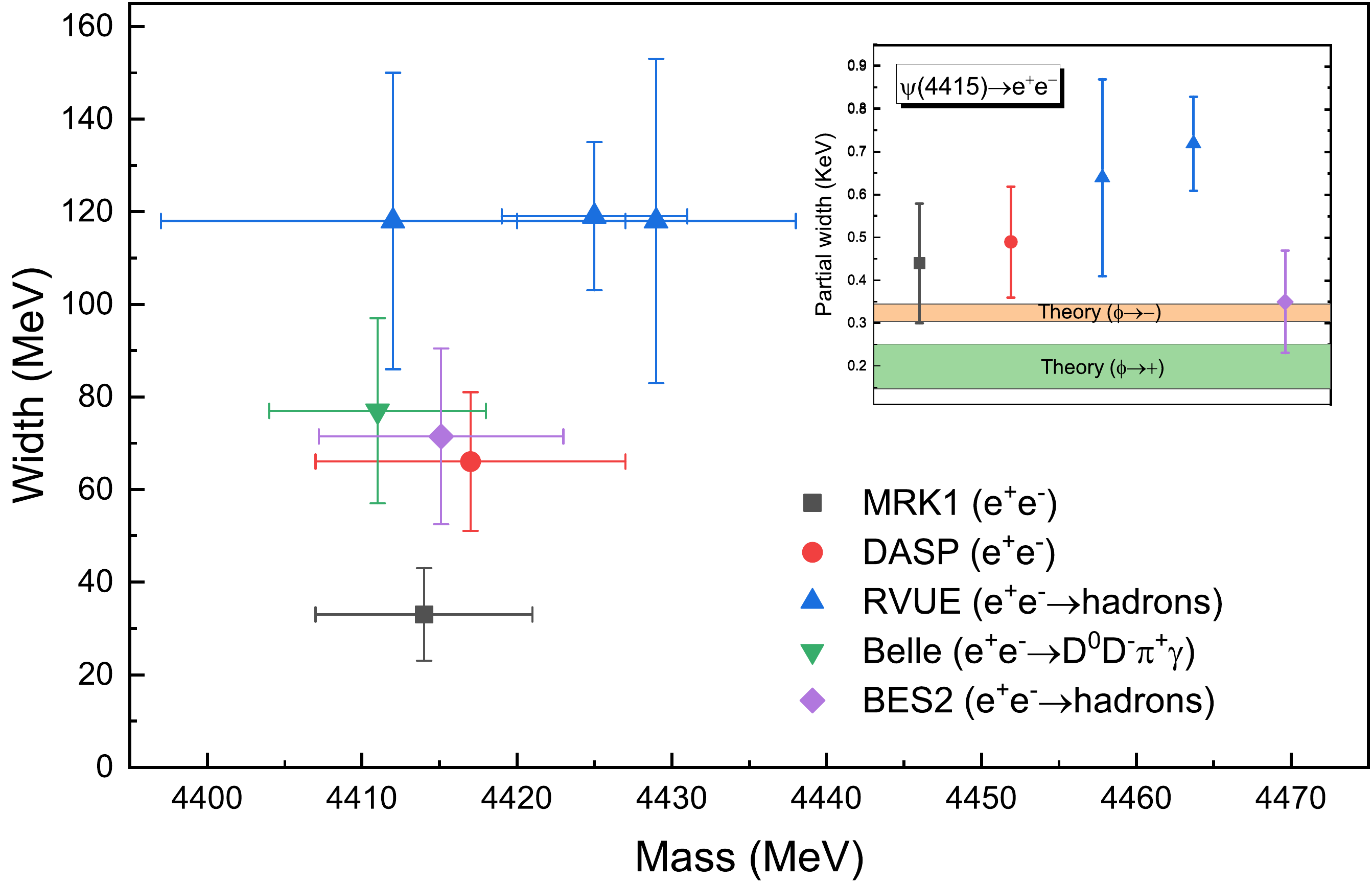}}
\caption{ The experimental resonance parameters and di-lepton width of $\psi(4415)$ from measurements of MARK1 \cite{Siegrist:1976br}, DASP \cite{Brandelik:1978ei}, RVUE \cite{Seth:2004py,Mo:2010bw}, Belle \cite{Pakhlova:2007fq}, and BESII Collaboration \cite{Ablikim:2007gd}.  \label{4415exp}}
\end{center}
\end{figure}

Under the $5S$-$4D$ mixing scheme, we illustrate the decay behaviors of $\psi(4415)$ and $\psi(4500)$ dependent on the mixing angle $\phi$ in Fig. \ref{5s4ddecay}. We notice that within the allowed range of $\phi$,
the theoretically obtained total decay width of $\psi(4415)$ is smaller than the average width value  ($62\pm20$ MeV) of $\psi(4415)$ collected in PDG \cite{Tanabashi:2018oca}.
Although 43 years have passed, the resonance parameters were not established well since the results from different experimental groups are very different. This can be seen in Fig. \ref{4415exp}, and most of the results are from inclusive processes of the $e^+e^-$ annihilation. Thus, we cannot test the assignment of $\psi(4415)$ as a $5S$-$4D$ mixing state by the present experimental width of $\psi(4415)$. Considering this situation, we strongly suggest to carry out the precise measurement of the resonance parameters of $\psi(4415)$ especially by exclusive processes (open-charm and hidden-charm channels) of the $e^+e^-$ annihilation, which will be an important task left for the experimentalist.

The dominant decay modes of the $\psi(4415)$ state are predicted to be $D^*D^*$, $DD_1(2420)$, and $DD_1(2430)$, and the corresponding branching ratios as a function of a mixing angle are shown in Fig. \ref{5s4ddecay}.
This means that the decay chains $\psi(4415) \to DD_1(2420)/DD_1(2430) \to DD^*\pi$ are allowed. The latest measurements of $e^+e^- \to D^0D^{*-}\pi^+$ by BESIII \cite{Ablikim:2018vxx} indeed indicate a possible signal near 4.42 GeV as shown in Fig. \ref{Opencharm2} (b).
In 2009, the BaBar Collaboration released two ratios \cite{Aubert:2009aq}
\begin{eqnarray}
\frac{\Gamma(\psi(4415)\to D\bar{D})}{\Gamma(\psi(4415) \to D^*\bar{D}^*)}=0.14\pm0.12\pm0.03, \\
\frac{\Gamma(\psi(4415)\to D^*\bar{D} + c.c.)}{\Gamma(\psi(4415)\to D^*\bar{D}^*)}=0.17\pm0.25\pm0.03,
\end{eqnarray}
which show the decay width of $\psi(4415)\to D^*\bar{D}^*$ is much larger than those of other two decay channels. This experimental result is consistent with our calculation for $\psi(4415)$.
Due to large errors, the above two experimental data cannot be applied to distinguish positive and negative mixing angles. Thus, more accurate measurements are still necessary.
In Table \ref{sdmixing}, we list the typical partial decay widths of the open-charm decay channels of $\psi(4415)$ and $\psi(4500)$ when taking typical $\phi=+30^\circ$ and $\phi=-30^\circ$.

Additionally, we need to point out that the $\psi(4415)$ decays into a pair of S-wave charmed-strange mesons are not obvious, which means that it is not an easy task to find a $\psi(4415)$ signal in the $e^+e^-\to D_{s}^{(*)}D_{s}^{(*)}$ processes. In Ref. \cite{delAmoSanchez:2010aa,Pakhlova:2010ek}, $e^+e^-\to D_s^{*+}D_s^{*-}$ and $e^+e^-\to D_s^{*+}D_s^{-}$
were analyzed, where $\psi(4415)\to D_s^{*+}D_s^{*-}$ was observed, but $\psi(4415)\to D_s^{*+}D_s^-$ was not seen, which are in accord with our results for the $D_sD^*_s$ and $D_s^*\bar{D}_s^*$ in Table \ref{sdmixing}.

Another interesting decay mode is $DD_2^*(2460)$, whose establishment has been experimentally achieved in  the cross section measurements of the process $e^+e^-\rightarrow D^0D^-\pi^+$ by Belle \cite{Pakhlova:2007fq}. As shown in Fig. \ref{Opencharm2} (a), they confirmed the existence of $\psi(4415)$, and released the peak cross section $\sigma(e^+e^-\to \psi(4415) \rightarrow D\bar{D}_2^*(2460) \rightarrow D^0D^-\pi^+ )=0.74\pm0.17\pm0.08 $ nb \cite{Pakhlova:2007fq}. Generally speaking, the branching ratio of $\psi(4415)\to D\bar{D}_2^*(2460)$ may be extracted from the above experimental data, and this depends on the input of mass, total decay width, and di-lepton width of $\psi(4415)$. We must note that the experimental resonance parameters of $\psi(4415)$ are quite inconsistent among several experiments as shown in Fig. \ref{4415exp}. Thus, we cannot directly compare the result of $\mathcal{B}(\psi(4415)\rightarrow DD_2^*(2460))$ between experiment and our prediction. Our theoretical result for $\mathcal{B}(\psi(4415)\to DD_2^*(2460))$ is $1.8\sim10.6 \%$ and $0.12\sim3.1\%$ in the negative and positive angles, respectively, which do not contradict the experimental observation.

We finally discuss the di-lepton width of $\psi(4415)$, which is the last remaining and available experimental information. Theoretically, the lepton width of a charmonium is proportional to a value of a resonance wave function at the origin. Our estimate gives $\Gamma(\psi(4415)\to e^+e^-)=0.303\sim0.344 $ or $0.147\sim0.251$ keV in the mixture scheme with negative and positive angles, respectively. Similar to the measured resonance parameters of $\psi(4415)$, the experimental differences of the di-lepton widths can be easily seen in Fig. \ref{4415exp}. Here, our di-lepton width can meet the measured values of $0.35\pm0.12$ and $0.44\pm0.14$ keV from BESII \cite{Ablikim:2007gd} and MARK1 \cite{Siegrist:1976br} within the experimental error range, respectively. Therefore, more precise measurements on $\psi(4415)$ are important to test our assignment of the $5S$-$4D$ mixing state.


We further predict the decay properties of the charmonium partner $\psi(4500)$ of $\psi(4415)$, whose total and partial widths of open-charm channels by varying the mixing angle are shown in Fig. \ref{5s4ddecay}. In the positive and negative mixing schemes, the total widths of $36\sim45$ MeV and $30\sim41$ MeV for $\psi(4500)$ are obtained, respectively. The dominant decay channels of $\psi(4500)$ are $DD_1(2430)$, $DD$, $D^*D^*$, $DD(2550)$, and $DD_1(2420)$. Additionally, the $DD^*_2(2460)$ and $DD^*$ are also main decay modes when mixing angle $\phi>0$, while the $D^*D_0^*(2400)$ is not negligible in $\phi<0$.
Their corresponding typical partial decay widths including both positive and negative mixing schemes are listed in Table \ref{sdmixing}. As for $\psi(4500)$, we can see that the decay modes $DD_1(2420)$, $DD_1(2430)$, $D^*D_0^*(2400)$, and $DD(2550)$ begin to become important. In addition to the two-body decay modes $DD$ and $D^*D^*$, the precise measurement of the three-body decay channel $DD^*\pi$ is also recommended in searching for the predicted $\psi(4500)$ in future. The lepton annihilation width of $\psi(4500)$ can also be predicted in $5S$-$4D$ mixing scheme, which is $2.25\times10^{-3}\sim0.0502$ and $0.0913\sim0.189$ KeV for the negative and positive angles, respectively. A such small lepton width compared with $\psi(4415)$, of course, causes the difficulty in searching for $\psi(4500)$ in the electron-positron collider.



\section{Hint of the predicted $\psi(4380)$ existing in $\mathbf{e^+e^- \to \psi(3686)\pi^+\pi^-}$ data}\label{sec3}

It is interesting to notice that almost all the vector charmonium-like states observed in the electron-positron annihilation process were observed in the hidden charm channels, such as $Y(4260)$, $Y(4220)$, and $Y(4320)$ in the $\pi^+ \pi^- J/\psi$ mode, $Y(4360)$ an $Y(4660)$ in the $\pi^+ \pi^- \psi(2S)$ mode, since the final states of these hidden charm decay modes are easier to be detected or reconstructed.   Therefore, a search for higher excited charmonia in the hidden charm processes will be interesting. However, as for higher excited charmonia, their mass splitting and their width are of the same order, thus the interferences between these resonances will become important. In Ref. \cite{Chen:2017uof}, the authors suggest that the experimental cross sections for $e^+e^- \to J/\psi \pi^+\pi^-$ and $e^+e^- \to h_c \pi^+\pi^- $ reported by the BESIII Collaboration can be reproduced by considering the contributions from three charmonium resonances $\psi(4160)$, $Y(4220)$, and $\psi(4415)$ and interferences with a nonresonance background, which is a kind of the Fano-like interference. Since such an interference effect is a general quantum phenomenon, it has been applied to atomic and nuclear physics a long time ago to understand experimental data \cite{fano-atomic,Orrigo:2006rd}. Specifically, the peak position of a genuine eigenstate is shifted by interferencing with the continuum via the Fano Hamiltonian and the corresponding Breit-Wigner distribution will be asymmetrically distorted  \cite{Fano:1961zz}. The application of the Fano interference effect can explain why two well established charmonium $\psi(4160)$ and $\psi(4415)$ have no obvious signals in the cross sections for $e^+ e^- \to \pi^+ \pi^- J/\psi$ and $e^+ e^- \to \pi^+ \pi^- h_c$ \cite{BESIII:2016adj, Ablikim:2016qzw}. Similarly, we can extend such a kind of analysis to the cross sections for $e^+e^- \to \psi(3686)\pi^+\pi^-$ in the present work.

Recently, the BESIII Collaboration reported their precise measurements of the cross sections for $e^+e^- \to \psi(3686) \pi^+\pi^-$ process \cite{Ablikim:2017oaf}, which provides us a good chance to revise whether there are more potential structures other than $\psi(4160)$ and $\psi(4415)$ as in Ref. \cite{Chen:2015bft}. In addition, it may provide an evidence of the predicted $\psi(4380)$ in the hidden-charm channel of $\psi(3686) \pi^+\pi^-$. In the Fano interference frame work, we firstly introduce an amplitude of a continuum background, which can be phenomenologically parameterized as,
\begin{eqnarray}
\mathcal{M}_{\rm NoR}=gu^2e^{-au^2}
\end{eqnarray}
with $u=\sqrt{s}-\sum_f m_f$ being the available kinetic energy, where $\sum_f$ is the sum of masses of final states. In the nonresonance amplitude, two phenomenological parameters $a$ and $g$ are introduced,  which are obviously related to non-perturbative QCD, and thus cannot be estimated from the first principle.

The genuine resonance contribution is described by a phase space corrected Breit-Wigner distribution, which is
\begin{eqnarray}
M_R(\psi)&=&\frac{\sqrt{12\pi\Gamma^{e^+e^-}_{\psi}\times \mathcal{B}(\psi\to\pi^+\pi^-\psi(3686))\Gamma_{\psi}}}{s-m_{\psi}^2+im_{\psi}\Gamma_{\psi}} \nonumber \\
&&\times\sqrt{\frac{\Phi_{2\to3}(s)}{\Phi_{2\to3}(m^2_{\psi})}},
\end{eqnarray}
where $\Phi_{2\to3}$ denotes the phase space of $e^+e^-\to\pi^+\pi^-\psi(3686)$ and $\psi$ is the intermediate vector charmonium. Here, the product of the electronic annihilation width $\Gamma^{e^+e^-}_{\psi}$ and branching ratio $\mathcal{B}(\psi\to\pi^+\pi^-\psi(3686))$ is treated as a free parameter $\mathcal{R}_{\psi}$. The total amplitude is the coherent sum of the nonresonance and resonance amplitudes, which is
\begin{eqnarray}
\mathcal{M}^{\rm Total}=\mathcal{M}_{\rm NoR}+\sum_k e^{i\phi_k} \mathcal{M}_R(\psi_k),
\end{eqnarray}
where $\phi_k$ is the phase angle between the continuum and the $k$-th intermediate resonance contribution.

\begin{figure}[htbp]
\begin{center}
\scalebox{0.41}{\includegraphics{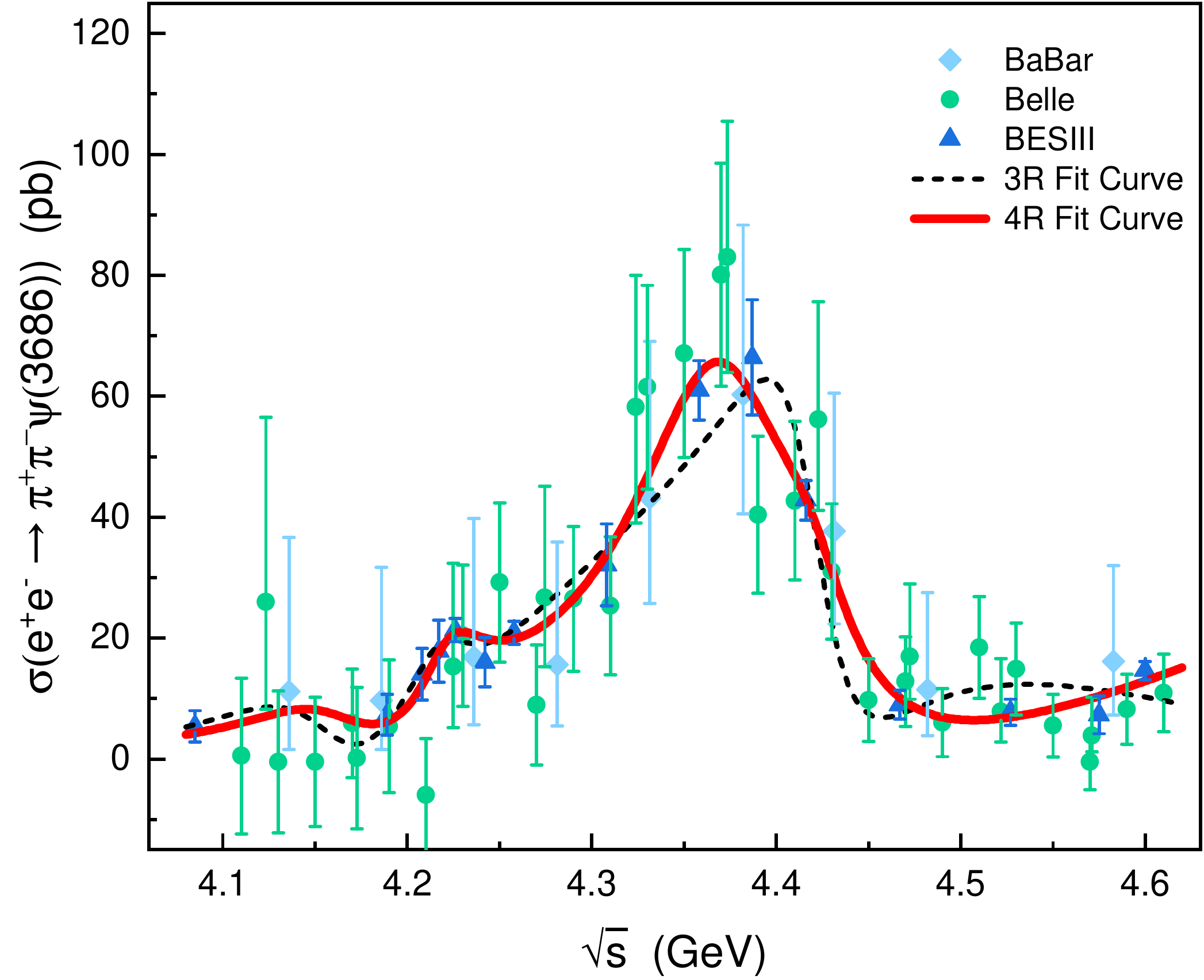}} \caption{ The fit to the cross section for the $e^+e^- \to \psi(2S)\pi^+\pi^-$ reaction in the Fano-like interference picture under $3R$ and $4R$ fit schemes. Here, the data of BaBar \cite{Aubert:2007zz}, Belle \cite{Wang:2007ea,Wang:2014hta} and latest BESIII \cite{Ablikim:2017oaf} are included.
\label{y4360}}
 \end{center}
\end{figure}

It is worth mentioning that $Y(4220)$ has been observed in the recent experimental data of $e^+e^-\to\pi^+\pi^-\psi(3686)$ from the BESIII Collaboration \cite{Ablikim:2017oaf}. So, we first fit the cross sections for $e^+ e^- \to \pi^+ \pi^- \psi(3686)$ with a nonresonance continuum and three genuine resonances, which are $\psi(4160)$, $Y(4220)$, and $\psi(4415)$.  We set the masses and widths of all the involved resonances to be the average values of PDG \cite{Tanabashi:2018oca}. The fitted results and corresponding parameters are shown in Fig. \ref{y4360} (black dashed curve) and Table \ref{fanopara}, respectively. It is interesting to notice that  most of the experimental data can be reproduced in $3R$ scenario with $\chi^2/d.o.f=1.22$. In particular, the enhancement signal of $Y(4220)$ is very clear.

The $3R$ scenario can reproduce most of the experimental data, and it should also be mentioned that the data from BESIII Collaboration obviously show the peak near 4.36 GeV in the fitted curve. This fact inspires us to propose an improved scheme, i.e., $4R$ fit scheme, where we consider an additional unknown $Y$ state with free mass and width to interfere with the background and other resonances contributions.  As  shown in Fig. \ref{y4360} (red solid curve), the experimental data can be perfectly reproduced in a $4R$ fit scheme, which is also reflected on an improved $\chi^2/d.o.f=0.748$. The resonance parameters of the $Y$ state are fitted to be
\begin{eqnarray*}
m&=&4374\pm13\, {\rm MeV},\\
\Gamma&=&106\pm29\, {\rm MeV},
\end{eqnarray*}
which are consistent with our predicted $\psi(4380)$ state. The above results indicate a structure near 4.37 GeV should exist and it cannot be simply described by the interferences from three resonances $\psi(4160)$, $Y(4220)$, and $\psi(4415)$ and continuum contribution. In other words, this conclusion shows a strong evidence of the existence of $\psi(4380)$ dominated by the $3D$-wave component in the hidden charm decay channel. At last, all of the puzzles are well resolved under our proposed theoretical picture, prompting us to have great confidence to believe that two longtime missing states $\psi(4S)$ and $\psi(3D)$ in the vector charmonium family could be experimentally established in the near future.

\begin{table}[htbp]
\renewcommand\arraystretch{1.1}
\caption{ The parameters obtained by fitting the cross section for $e^+e^- \to \psi(2S) \pi^+\pi^-$ \cite{Ablikim:2017oaf,Aubert:2007zz,Wang:2007ea,Wang:2014hta}. \label{fanopara}}
\begin{center}
{\tabcolsep0.10in
\begin{tabular}{ccc}
\toprule[1pt]\toprule[1pt]
  Parameters & 3R Fit & 4R Fit \\
\midrule[1pt]
$g$ (MeV$^{-1}$)                     & $4.13 \pm 0.32$                &$0.79\pm0.12$                                 \\
$a$ (GeV$^{-2}$)                  &    $5.11\pm0.20$                &   $0.62\pm0.50$                                \\
$\mathcal{R}_{\psi(4160)}$ (eV)             &$2.59 \pm 0.86$                &$4.75\pm1.22$                        \\
 $\phi_1$ (rad)                         &$4.54\pm0.15$                &$5.33\pm0.25$                          \\
 $\mathcal{R}_{Y(4220)}$ (eV)          &$0.15\pm 0.09$                &$5.67\pm0.61$                          \\
 $\phi_2$ (rad)                       &$1.79\pm0.35$                & $3.84\pm0.26$                                   \\
 $\mathcal{R}_{\psi(4415)}$ (eV)         &$2.34\pm0.46$                &$0.30\pm0.45$                          \\
 $\phi_3$ (rad)                          &$3.39\pm0.10$              &$2.68\pm0.98$               \\
 $m_Y$ (MeV)                      &$\cdots$                          &$4374\pm13$                            \\
 $\Gamma_Y$ (MeV)            &$\cdots$                              &$106\pm29$                      \\
  $\mathcal{R}_Y$ (eV)            &$\cdots$                    &$6.61\pm2.91$              \\
   $\phi_4$ (rad)          &$\cdots$                           &$2.40\pm0.46$                     \\

\midrule[1pt]
 $\chi^2/d.o.f$					   & 1.22	      & 0.748        \\

\bottomrule[1pt]\bottomrule[1pt]
\end{tabular}
}
\end{center}
\end{table}

\section{Summary}\label{sec4}
The observation of $J/\psi$ particle in 1974 opens a new era of particle physics \cite{Aubert:1974js,Augustin:1974xw}. Since then, more and more charmonia have been reported by experiments, which construct the main body of the present $c\bar{c}$ meson spectrum as listed in PDG  \cite{Tanabashi:2018oca}. Although the $J/\psi$ family has become abundant with the effort made by experimentalists, the $J/\psi$ family is far
from being well established. In the past 15 years, the observations of a series of charmoniumlike states have brought us a new chance and challenge to study $c\bar{c}$ meson spectrum \cite{Liu:2013waa,Chen:2016qju}. It is obvious that it is also a good opportunity for hadron physics.

In this work, we have focused on the updated data of charmoniumlike $Y$ states from the $e^+e^-$ annihilations, and have further revealed that $Y(4220)$ observed in the $e^+e^- \to J/\psi\pi^+\pi^-$ processes is an important scaling point when constructing higher charmonia. Here, $Y(4220)$ has been established as a charmonium under $4S$-$3D$ mixing scheme, and further theoretical prediction of its decay behaviors has been given, which provides valuable information to test this scenario. What is more important is that we have also predicted the existence of the charmonium partner $\psi(4380)$ of $Y(4220)$. According to our calculation, we have obtained its resonance parameters and partial open-charm decay widths. Furthermore, we have also discussed how to identify
the predicted $\psi(4380)$ by the present data of open-charm and hidden-charm decay channels. Especially, we have analyzed the latest experimental data of  $e^+e^-\to \psi(3686)\pi^+\pi^-$ measured by BESIII \cite{Ablikim:2017oaf} by combining with the Fano interference picture, where the possible evidence of $\psi(4380)$ has been found. Hence, we suggest future experiments like BESIII and Belle II to hunt for $\psi(4380)$, which not only tests this charmonium assignment to $Y(4220)$, but pushes experimental progress on charmonium or charmoniumlike states.

When finishing the study, we have to face another crucial issue, i.e., how to settle the charmonium $\psi(4415)$. In this work, we have investigated $\psi(4415)$ under $5S$-$4D$ mixing scheme, and have found that the obtained results do not contradict with the experimental data of $\psi(4415)$.
If carefully checking the present experimental information listed in PDG \cite{Tanabashi:2018oca}, we notice that the precision of data is not enough since even the first observation of $\psi(4415)$ has passed 42 years \cite{Siegrist:1976br}. Therefore, further experimental studies on $\psi(4415)$ are strongly encouraged, especially at BESIII and Belle II. As a charmonium partner of $\psi(4415)$, a missing charmonium $\psi(4500)$ has been predicted in this work. The search for it will be an interesting research issue.

We hope that our theoretical studies presented here can play an important role in constructing the $J/\psi$ meson spectrum, especially higher charmonia. More experimental and theoretical joint efforts on this topic will be necessary in forthcoming years.


\section*{Acknowledgments}

This work is partly supported by the China National Funds for Distinguished Young Scientists under Grant No. 11825503, the National Natural Science Foundation of China under Grant No. 11775050, National Program for Support of Top-notch Young Professionals, and the Fundamental Research Funds for the Central Universities.

{\bf Note added}: When we are writing out the present work, we have noticed a recent result from BESIII \cite{Ablikim:2019apl}. By analyzing the data of the cross section of $e^+e^-\to \omega\chi_{c0}$ from
$\sqrt{s}=4.178$ to $4.378$ GeV, BESIIII has confirmed the existence of a narrow structure $Y(4220)$ at 4.2 GeV.  Especially, BESIII has also extracted the angular distribution of $e^+e^-\to \omega\chi_{c0}$, which shows that there exists the evidence for a combination of $S$ and $D$-wave contribution in the $Y (4220) \to \omega\chi_{c0}$ \cite{Ablikim:2019apl}. This updated measurement of  $e^+e^-\to \omega\chi_{c0}$ supports our $4S$-$3D$ mixing scheme for $Y(4220)$.

\end{document}